\documentclass[fleqn,usenatbib]{mnras}

\usepackage[T1]{fontenc}
\usepackage{color}
\usepackage[normalem]{ulem}

\DeclareRobustCommand{\VAN}[3]{#2}
\let\VANthebibliography\thebibliography
\def\thebibliography{\DeclareRobustCommand{\VAN}[3]{##3}\VANthebibliography}

\usepackage{graphicx}	% Including figure files
\usepackage{amsmath}	% Advanced maths commands
\usepackage{bm}
\usepackage{amssymb}	% Extra maths symbols
\usepackage[noline, boxed,commentsnumbered]{algorithm2e}
\SetAlCapSkip{1em}
\usepackage{booktabs}
\usepackage{pdflscape}
\usepackage{subcaption}
\usepackage{multicol}
\usepackage{tabularx}

\newcommand\given[1][]{\:#1\vert\:}

\newcommand{\tabhead}[1]{\textbf{#1}}

\title[Unsupervised Machine Learning]{Unsupervised Machine Learning for the Classification of Astrophysical X-ray Sources}

\author[Pérez-Díaz et al.]{Víctor Samuel Pérez-Díaz$^{1,2}$\thanks{E-mail: samuelperez.di@gmail.com},
Juan Rafael Martínez-Galarza$^{1}$,
Alexander Caicedo$^{3, 4}$ and
\newauthor
Raffaele D'Abrusco$^{1}$
\\
$^{1}$Center for Astrophysics | Harvard \& Smithsonian, 60 Garden Street, Cambridge, MA 02138, USA\\
$^{2}$School of Engineering, Science and Technology, Universidad del Rosario, Cll. 12C No. 6-25, Bogotá, Colombia\\
$^{3}$Department of Electronics Engineering, Pontificia Universidad Javeriana, Cra. 7 No. 40-62, Bogotá, Colombia \\
$^{4}$Ressolve, Cra. 42 \# 5 Sur - 145, Medellín, Colombia
}

\date{Accepted XXX. Received YYY; in original form ZZZ}

\pubyear{2022}

\usepackage{newtxtext,newtxmath}

\begin{document}
\label{firstpage}
\pagerange{\pageref{firstpage}--\pageref{lastpage}}
\maketitle

\begin{abstract}
The automatic classification of X-ray detections is a necessary step in extracting astrophysical information from compiled catalogs of astrophysical sources. Classification is useful for the study of individual objects, statistics for population studies, as well as for anomaly detection, i.e., the identification of new unexplored phenomena, including transients and spectrally extreme sources. Despite the importance of this task, classification remains challenging in X-ray astronomy due to the lack of optical counterparts and representative training sets. We develop an alternative methodology that employs an unsupervised machine learning approach to provide probabilistic classes to Chandra Source Catalog sources with a limited number of labeled sources, and without ancillary information from optical and infrared catalogs. We provide a catalog of probabilistic classes for 8,756 sources, comprising a total of 14,507 detections, and demonstrate the success of the method at identifying emission from young stellar objects, as well as distinguishing between small-scale and large-scale compact accretors with a significant level of confidence. We investigate the consistency between the distribution of features among classified objects and well-established astrophysical hypotheses such as the unified AGN model. This provides interpretability to the probabilistic classifier. Code and tables are available publicly through GitHub. We provide a web playground for readers to explore our final classification at \url{https://umlcaxs-playground.streamlit.app}.
\end{abstract}


%%%%%%%%%%%%%%%%%%%%%%%%%%%%%%%%%%%%%%%%%%%%%%%%%%

%%%%%%%%%%%%%%%%% BODY OF PAPER %%%%%%%%%%%%%%%%%%

\section{Introduction}

X-ray astronomy is embarking on a promising era for discovery, facilitated by cutting-edge missions like \textit{eROSITA}. This mission, which scans the entire X-ray sky, has already made its initial data release accessible to the scientific community \citep{predehl2021erosita, merloni2012erosita}.  Prospective missions such as \textit{Athena}  and \textit{Lynx} offer enhanced capabilities. Specifically, \textit{Athena} combines the sensitivity of \textit{Chandra} with the expansive coverage area of \textit{eROSITA} \citep{nandra2013hot}. Meanwhile, \textit{Lynx} plans to surpass \textit{Chandra} in terms of sensitivity and resolution \citep{gaskin2019lynx}. These missions will provide a wealth of new data and detections, catalyzing the identification of previously unrecognized sources.

The \textit{Chandra X-ray Observatory} is one of the NASA's great observatories and its flagship mission for X-ray astronomy. Since it was launched to space in 1999, the telescope has observed several sources with two instruments, the Advanced CCD Imaging Spectrometer (ACIS) and the High Resolution Camera (HRC) \citep{chandra_20_book}. Accreting black holes in the center of galaxies, supernova remnants, X-ray binaries, and young, rapidly rotating magnetic stars are some of the most common targets that Chandra has identified. As one of NASA's Great Observatories, Chandra has contributed to numerous groundbreaking discoveries in high-energy astrophysics over its 23-year lifespan. The Chandra Source Catalog (CSC) collects the X-ray sources detected by the Chandra X-ray Observatory through its history \citep{Evans2010}. It represents a fertile background for discovery, since many of this sources have not been studied in detail. A significant proportion of these sources, which range from young stellar objects (YSO) and binary systems (XB) to distant active galaxies housing supermassive black holes (AGN), remain largely unexplored. Furthermore, Chandra's data contains traces of exceptional phenomena like extrasolar planet transits, tidal disruption events, and compact object mergers. Despite the potential scientific wealth within Chandra's data, only a fraction of CSC sources have been classified. In order to conduct a thorough investigation of CSC sources, and to gear up for forthcoming large-scale X-ray surveys, we need to classify as many catalog sources as possible.

As larger astronomical surveys become available, researchers are increasingly adopting more sophisticated statistical models, including machine learning and data science methods, for classification tasks. A comparison of various supervised and unsupervised learning methods for the statistical identification of XMM-Newton sources was presented in \cite{pineau2010comparison}. This study employed a probabilistic cross-correlation of the XMM-Newton Serendipitous Source Catalog (2XMMi) with catalogs such as Sloan Digital Sky Survey (SDSS) DR7 and Two Micron All-Sky Survey (2MASS). They compared algorithms like k-Nearest Neighbors, Mean Shifts, Kernel Density Classification, Learning Vector Quantization, and Support Vector Machines.

In \cite{lo2014automatic}, an automatic classification method for variable X-ray sources in 2XMMi-DR2 using Random Forest was presented. The authors achieved approximately 97\% accuracy for a 7-class dataset. In \cite{farrell2015autoclassification}, a catalog of variable sources in 3XMM was classified using Random Forest, training the classifier on manually classified variable stars from 2XMMi-DR2 and obtaining approximately 92\% accuracy.

As interest in this research field continues to grow, unsupervised learning techniques have gained prominence in recent years. In \cite{rostami2019classification}, an automated machine learning tool for classifying extra-galactic X-ray sources using multi-wavelength data, particularly from the Hubble Space Telescope, was proposed. In \cite{ansariagnellogall}, a probabilistic assignment was performed using mixture density networks (MDN) and infinite Gaussian mixture models to classify objects in the dataset as stars, galaxies, or quasars, achieving a 94\% accurate split. The training data consisted of magnitudes from SDSSDR15 and the Wide-field Infrared Survey Explorer (WISE).

In the same line, in \cite{loganfotopoulou}, an alternative unsupervised machine learning method for separating stars, galaxies, and QSOs using photometric data was presented. This approach employed Hierarchical Density-Based Spatial Clustering of Applications with Noise (HDBSCAN) to identify different classes in a multidimensional color space. Using a constructed dataset of approximately 50,000 spectroscopically labeled objects, the authors achieved F1 scores of ${>}0.92$ for star, galaxy, and QSO classes.

In recent years, significant efforts have been made to improve the classification of X-ray sources using supervised machine learning methods. Notably, \cite{yang2022classifying} developed a novel approach for classifying CSC version 2 (CSC2) sources.  They successfully classified $66369$ CSC2 sources (21\% of the entire catalog) and performed focused case studies, demonstrating the versatility of their machine learning approach. Additionally, they provided insights into the biases, limitations, and bottlenecks encountered in these types of studies. Similarly, \cite{kumaran2023automated} aimed to classify point X-ray sources in the Chandra Source Catalog 2.0 into categories such as active galactic nuclei (AGN), X-ray emitting stars, young stellar objects (YSOs), high-mass X-ray binaries (HMXBs), low-mass X-ray binaries (LMXBs), ultra-luminous X-ray sources (ULXs), cataclysmic variables (CVs), and pulsars. Using multi-wavelength features from various catalogs, including CSC2, they used the Light Gradient Boosted Machine algorithm and achieved scores of ${\geq}91$\% in precision, recall, and Matthew's Correlation coefficient.

Despite the recent efforts, supervised classification of X-ray sources remains a complex endeavor. Building a reliably labeled training set often involves manual review of the existing literature, which can sometimes contain ambiguous results. Moreover, a large number of these sources are missing optical or infrared data that could provide additional insights into their physical nature. Thus, unsupervised learning offers a distinct advantage as it eliminates the need for a pre-labeled training set. It can also reveal hidden patterns in the data that may not be immediately apparent. In this study, we introduce an unsupervised learning approach designed to classify as many CSC sources as possible, leveraging solely available X-ray data. By clustering the source observations by their similarities, and then associating these clusters with objects previously classified spectroscopically, we propose a new methodology to provide probabilistic classification for a numerous amount of sources. Furthermore, we offer not only a classification, but also a comprehensive discussion on the astrophysical implications arising from the observations made at each stage of the pipeline. We evaluate the strenghts and limitations of the method. 

In Section \ref{sec:data}, we describe the primary dataset used in this work, the Chandra Source Catalog, detailing the query and preprocessing steps required to make the data suitable for analysis with unsupervised learning methods. We also discuss the feature selection process and final properties used. In Section \ref{sec:methods}, we explain the methods employed, including Gaussian Mixtures and the development of a probabilistic classification algorithm based on the Mahalanobis Distance. We also introduce Soft/Hard voting classifiers that were used for the final output of this study. This section includes an overview of the techniques, a detailed explanation of the algorithms, relevant parameter choices, and a validation approach for the method. In Section \ref{section:results}, we provide a comprehensive exploration of results for each step of the algorithm. We report per-detection classifications, final master classes for 8,756 sources, and significant trends or patterns observed during the analysis. In Section \ref{sec:discussion}, we discuss the implications of our results, including comparisons with previous work and catalogs. We analyze the astrophysical nature revealed by the data and evaluate the possible limitations of our study, addressing potential biases and sources of uncertainty. We also suggest directions for future research. Finally, in Section \ref{sec:conclusions}, we summarize the conclusions of the work, highlighting the main outcomes and their relevance.

\section{Data}
\label{sec:data}
In this section, we describe the data set used in the work at hand, and the properties chosen for the subsequent analysis. Our main data consists of Chandra Source Catalog version 2 per-detection registers.

\subsection{The Chandra Source Catalog}

The Chandra Source Catalog (CSC) collects and presents summarized properties for the X-ray sources detected by the Chandra X-ray Observatory through its history. Version 2.0 (CSC2), which we use here, is the second major release of the catalog, including properties for $317167$ X-ray sources in the sky. Properties include source photometry (brightness), spectroscopy (energy), and variability (change over time).

The Chandra Source Catalog includes properties for $928280$ source detections, which were detected in $10382$ Chandra observations until $2014$. The catalog presents information through $1700$ columns of tabular data, distributed across five energy bands (broad, hard, medium, soft, and ultra-soft) for the Advanced CCD Imaging Spectrometer (ACIS), and a single band for the High Resolution Camera (HRC) in the wide spectrum. Chandra has been observing the universe in the 0.5-8 KeV band \citep{chandra_20_book}. The master source properties present summary measurements of stacked detections. Properties for detected sources are measured for both the stacked detection and the individual observations in all of the bands \citep{chandra_20_book}. 

This study utilizes the individual properties measured for detections in the Chandra Source Catalog. Our initial classification is performed on a per-observation basis, before we evaluate the most probable class for each master source. This approach takes into account that CSC sources may have multiple detections. A subset of the CSC2 per-observation detection table was extracted using CSCView\footnote{\url{http://cda.cfa.harvard.edu/cscview/}}, with selection criteria restricting source detections to those with a broad band flux significance greater than 5 ($\texttt{flux\_significance\_b} > 5$), and non-null values for power-law gamma (\texttt{powlaw\_gamma}) and blackbody temperature (\texttt{bb\_kt}). These criteria aimed to ensure observations possessed sufficient significance to allow for fitting their spectra to a model, thereby enabling the extraction of statistical relationships from the data. This process yielded a table of $37878$ unique observational entries.

\subsection{Properties and preprocessing}

In this subsection, we discuss the properties used in our analysis. We also outline the preprocessing steps applied to the dataset. These steps include data cleaning, transformation, and normalization. We conducted an iterative exploration process for feature selection. We selected 12 properties, primarily involving variability and spectral measurements. Most properties are extracted directly from the per-observation level of the CSC, except for \texttt{var\_ratio\_*} and \texttt{var\_newq\_b}. These were computed by us to summarize source information from variability mean, standard deviation, min and max. Table \ref{tab:properties} lists the chosen properties and their descriptions.

\begin{table}
\caption{Properties selected for our analysis. Properties denoted with '\_*' represent three features for energy bands: \textit{b} (broad), \textit{h} (hard), \textit{s} (soft). \textit{Adapted from \href{https://cxc.cfa.harvard.edu/csc/}{CSC documentation webpage}.}}
\label{tab:properties}
\centering
\begin{tabularx}{0.5\textwidth}{|l|X|}
\hline
\tabhead{property name} & \tabhead{description} \\ \hline
hard\_hm & ACIS hard (2.0-7.0 keV) - medium (1.2-2.0 keV) energy band hardness ratio - basically the ratio between the hard and medium energy bands\\ \hline
hard\_hs & 	ACIS hard (2.0-7.0 keV) - soft (0.5-1.2 keV) energy band hardness ratio - basically the ratio between the hard and soft energy bands\\ \hline
hard\_ms & ACIS medium (1.2-2.0 keV) - soft (0.5-1.2 keV) energy band hardness ratio - basically the ratio between the medium and soft energy bands\\ \hline
bb\_kt & temperature (kT) of the best fitting absorbed black body model spectrum to the source region aperture PI spectrum - temperature of the object estimated by a black body model.\\ \hline
powlaw\_gamma  & photon index of the best fitting absorbed power-law model spectrum to the source region aperture\\ \hline 
var\_prob\_* & intra-observation Gregory-Loredo variability probability (highest value across all stacked observations) for each science energy band - variability probability in a single observation with Gregory-Loredo technique.\\ \hline
var\_ratio\_* & the ratio of flux variability mean value to its standard deviation
$$
\frac{\text{var\_mean\_*}} {\text{var\_sigma\_*}}
$$
\\ \hline
var\_newq\_b & proportion of the average of minimum and maximum count rates (i.e., data points in the light curve) during an observation relative to the mean count rate.
$$
\frac{\text{var\_max\_b + var\_min\_b}}{2 \cdot \text{var\_mean\_b}}
$$
\\
\hline
\end{tabularx}
\end{table}

Our initial step was to cleanse the original data, selecting rows with valid entries (not NaN) for the chosen properties. This reduced the dataset to $29655$ rows. Following this, we carried out either normalization exclusively or a combination of logarithmic transformation and normalization, depending on the distribution and range of the selected properties. Feature preprocessing was tailored to achieve scale uniformity, preventing large-scale features, like outliers, from dominating the learning process and facilitating faster convergence. Details of the chosen processing approach for each property are presented in Table \ref{tab:prop_lognorm}. When required, the log transformation is performed prior to normalization, feeding its result into the normalization process.  

\begin{table}
\caption{Preprocessing approach for each property. Energy bands are denoted by \textit{* = b, h, s} (broad, hard, and soft).}
\label{tab:prop_lognorm}
\centering
\begin{tabularx}{0.5\textwidth}{|X|c|c|}
\hline
\tabhead{property name} & \tabhead{log} & \tabhead{normalization}\\ \hline
hard\_hm &  &\\ \hline
hard\_hs & 	&\\ \hline
hard\_ms & &\\ \hline
bb\_kt & $\bullet$ & $\bullet$\\ \hline
powlaw\_gamma  & & $\bullet$\\ \hline 
var\_prob\_* & &\\ \hline
var\_ratio\_* & $\bullet$ & $\bullet$\\ \hline
var\_newq\_b & $\bullet$ & $\bullet$\\ \hline
\end{tabularx}
\end{table}

    \subsubsection{Normalization}
    
    We employ the MinMaxScaler method from the \texttt{scikit-learn} Python library \citep{scikit-learn} for data normalization. Using this, we scale each selected feature to the range $[0,1]$ based on its minimum and maximum values. The transformation can be represented as:
    
    \begin{equation}
    \label{eq:minmax}
        \mathbf{X}_{scaled} = \frac{\mathbf{X} - \mathbf{X}_{\text{min}}}{\mathbf{X}_{\text{max}} - \mathbf{X}_{\text{min}}}.
    \end{equation}

    In equation \ref{eq:minmax}, $\mathbf{X}$ denotes the original feature values, while $\mathbf{X}_{\text{min}}$ and $\mathbf{X}_{\text{max}}$ represent the minimum and maximum values of the feature, respectively. The resulting $\mathbf{X}_{scaled}$ maintains the relative relationships between the original values.
    
    \subsubsection{Log transformation}
    
    We apply a natural logarithm to each data point using the \texttt{numpy} library \citep{numpy}. To handle zero values, we add one-tenth of the non-zero minimum value of each property to the data before taking the logarithm:

    \begin{equation}
    \label{eq:logtransform}
    \begin{split}
    X_{\text{min}}^* &= \min\{x_i \in \mathbf{X} \mid x_i \neq 0, i = 1, \dots, n\} \\
    \mathbf{X}_{\text{log}} &= \log \left(\mathbf{X} + \frac{X_{\text{min}}^*}{10}\right)
    \end{split}
    \end{equation}

    In the equations above, $X_{\text{min}}^*$ represents the minimum non-zero value in the feature values $\mathbf{X}$. The log-transformed property is represented by $\mathbf{X}_{\text{log}}$.

\subsection{Feature selection}
\label{sec:results_1_1}
The efficacy of our proposed algorithm relies significantly on the selection of CSC properties used as features. Feature selection in unsupervised machine learning is an active field of research \citep{solorio2020review}. For this study, we prioritized features that enhance information gain, i.e, those that maximize the separation between clusters. While dimensionality reduction techniques are often used in feature selection, we chose to follow a different criteria to prioritize interpretability. Additionally, feature scales vary significantly, adding complexity to the task. Selected \textit{CSC} properties are described in Table \ref{tab:properties}, and they were chosen using the following criteria:

\begin{enumerate}
    \item We start from astrophysical domain knowledge. Features that are more likely to inform the classification are those related with the spectral and time-domain properties, because they are associated to specific astrophysical processes (e.g. accretion).
    
    \item Besides using astrophysically significant features, we also employ an empirical strategy. Here, we evaluate feature importance by the degree to which a given set of features yields distinct clusters. We do this by observing how the properties distribute in each cluster for various feature selections and combinations.
%, like those depicted in Figures \ref{fig:hard_hs_hist} and \ref{fig:var_prob_hist},
    \item Then, we choose properties based on maximizing the amount of information in clusters while simultaneously minimizing redundancies. By minimizing redundancy, we can reduce the number of features required to accurately classify sources, which may lead to simpler and more interpretable models. At the same time, maximizing the amount of information in clusters ensures that the clusters formed are distinct and informative. This approach addresses the challenge of determining the optimal number of astrophysical x-ray properties required for accurate source classification, and provides insights into the most important properties for distinguishing between different source types.
\end{enumerate}

\section{Methods}
\label{sec:methods}
In this section we present the unsupervised learning and cluster-wise classification techniques employed in our analysis. We use Gaussian Mixtures Models (GMMs), which allow dataset clustering based on the multi-dimensional distribution of data points. In this study, selected CSC column properties serve as the features for our analysis. We first run a GMM on those features to find initial clusters. We then cross-match our CSC sources with the SIMBAD database in order to extract any existing labels assigned to the observations. We assign probabilistic classes to each unlabelled observation inside each group. The assignment is accomplished by examining the distance between the unlabelled observation and the centroids of labeled observations (extracted from SIMBAD) that share the same class within the cluster. Based on the detection classifications, we assign a final class to the master source. The next subsections offer a more detailed account of each step in this algorithm.

\subsection{Gaussian Mixture Models and the \textit{EM} algorithm}
\label{subsec:methods_1}
Clusters are first identified using a Gaussian Mixture Model (\textit{GMM}) in the multi-dimensional space of the chosen \textit{CSC} properties. In order to understand the assumptions and limitations of this method, we first have to describe the basics of its functionality as a clustering technique. A Gaussian Mixture represents a linear combination of $K$ different Gaussian distributions \citep{bishop}

\begin{equation}
    \label{eq:eq1}
    p(\bm{x}) = \sum_{k=1}^K \pi_k \mathcal{N}(\bm{x} \given \bm{\mu}_k, \bm{\Sigma}_k)
\end{equation}

\noindent where $\pi_k$ are the \textit{mixture coefficients} or \textit{mixture weights}, and

\begin{equation}
    \label{eq:eq2}
    0 \leq \pi_k \leq 1, \;\;\; \sum_k \pi_k = 1.
\end{equation}

\noindent %thus satisfying the conditions that allow us to interpret $p(\bm{x})$ as a probability.%
The Gaussian distributions $\mathcal{N}(\bm{x} \given \bm{\mu}_k, \bm{\Sigma}_k)$ that compose the mixture are called \textit{components}, each having its own covariance $\bm{\Sigma}_k$ and mean $\bm{\mu}_k$. We use maximum likelihood estimation in order to adjust the parameters of the mixture of Gaussians to optimally represent clusters in the data. From \ref{eq:eq1}, the log likelihood function is given by

\begin{equation}
    \label{eq:eq3}
    \ln p(\bm{X} \given \bm{\pi}, \bm{\mu}, \bm{\Sigma})  = \sum_{n=1}^N \ln \sum_{k=1}^K \pi_k \mathcal{N}(\bm{x} \given \bm{\mu}_k, \bm{\Sigma}_k).
\end{equation}

In the case of a single Gaussian distribution, an analytical solution for the maximum likelihood can be obtained easily. Indeed, in the simplest case (one dimensional) it corresponds to the mean and variance of the data. However, maximizing the likelihood of a Gaussian Mixture is not an easy task, lacking a closed-form analytical solution. A broadly applicable method solution, which we used in this paper, is the \textit{expectation-maximization} algorithm, also abbreviated as \textit{EM} algorithm.

The \textit{EM} algorithm is a method for estimating the parameters that maximize the likelihood in unobserved latent variable dependent models. In this case, the latent variable is the cluster assignment of each data point. The \textit{EM} algorithm was initially proposed in \cite{em_dempster}. A complete review of the method could be found in \cite{mclachlan2007algorithm}.

The \textit{EM} algorithm consist of two main states, as the name suggests: the \textit{expectation} (\textit{E}) step and the \textit{maximization} (M) step. The initial parameters of the Gaussian Mixture could be chosen in different approaches. The most common one would use random initial parameters. However, a more convenient routine is to assume identity covariances in the Gaussians and perform \textit{K-means} clustering to find the means. A summary of the \textit{EM} algorithm steps is described in online Appendix A.

We have provided a general description of the \textit{GMM} and the \textit{EM} algorithm. For the mathematical details, we refer the reader to \cite{bishop, Deisenroth2020}. An analysis of how the log-likelihood increases in each of the steps of the \textit{EM} routine is presented in \cite{Neal1998}.

In this work, we use the \textit{GMM} architecture provided in the \textit{scikit-learn} library \citep{scikit-learn} for Python. There is no direct evidence supporting the assumption that our data come from a mixture of gaussians. However, \textit{GMM} is a flexible method capable of producing reliable results even when this assumption is not strictly met. We expect this minimizes the impact of the possible more complex distributions present in the data. The Gaussian Mixture serves as a foundational step in our classification process. Throughout our analysis, components and clusters will mean the same. The selection of hyperparameters for our study is explained in Section \ref{sub:hyperparameter}.

\subsection{Cluster-wise classification algorithm}
\label{sec:methods_2}

\subsubsection{Per-detection classification}
The initial clustering serves as a preliminary step towards classification. Given the complexity of astrophysical classes, they likely exceed the optimal number of clusters we will derive. This can be partly attributed to the potential overlap of various astrophysical classes within the feature space, particularly when relying solely on X-ray data. Consequently, a single cluster may encompass objects of multiple classes. However, it is expected that object detections within the same cluster will exhibit similar properties.

The \textit{SIMBAD Astronomical Database}, operated at CDS, Strasbourg, France, \citep{simbad_wengeretal} provides a relatively solid knowledge base source for a systematic class extraction \citep{simbadcategorization} \footnote{Object types (\texttt{otypes}) description URL:
\url{https://simbad.astro.unistra.fr/guide/otypes.htx}}. SIMBAD employs a hierarchical system for classifying objects, which relies on catalogue identifiers. SIMBAD provides a base for associating probabilistic classses to unclassified object detections. Initially, we cross-match our dataset with classified objects in the SIMBAD database, leading to a minority of classified objects per cluster. We then compute the distance between each unclassified object and the classes centroid within the cluster. These distances facilitate probabilistic assignment, providing an indication of the detection's likelihood of belonging to a given class.

The crossmatch was performed using the \textit{CDS Upload X-Match} functionality of \textit{TOPCAT} \citep{topcat_taylor}, which is an interface for the crossmatch service provided by CDS, Strasbourg. The match radius was set to ${\leq}1\arcsec$ and the Find mode was set to \textit{Each} to obtain a new table with the best match from the \textit{SIMBAD} database, if any, within the selected radius. In the absence of a match, the \textit{SIMBAD} columns in the table would be blank. We specifically looked for the \texttt{otype} property, which refers to the main object type assigned to the sources \citep{anaisetal}. It is expected that a significant number of sources would not have a matching entry in the \textit{SIMBAD} database or may have an ambiguous \texttt{otype} classification, such as peculiar emitters (Radio, IR, Red, Blue, UV, X, or gamma), assigned when no further information is available about the nature of the source.

We performed the classification of sources using the extracted labels and cluster information. To achieve this, we measured the similarity between the unlabeled and the labeled observations within a cluster using the \textit{Mahalanobis} distance metric. %This method takes into account the unique distribution of the classes within the cluster space, making it a suitable choice for our analysis.

The \textit{Mahalanobis Distance} was introduced in \cite{mahalanobis1936generalized}, as a distance metric of a point and a distribution. It is defined as:

\begin{equation}
    \label{eq:methods_eq6}
    D_P(\bm{x})= \sqrt{(\bm{x} - \bm{\mu})^T \Sigma^{-1}(\bm{x} - \bm{\mu})},
\end{equation}

where $\bm{\mu}$ and $\Sigma$ are the mean and covariance matrix of a probability distribution $P$ respectively, and $\bm{x}$ is a data point. 

%The use of the \textit{Mahalanobis} distance as the metric for our analysis was motivated by its ability to take into account in the evaluation of the distance within the dataset, specifically the distributions of each class group within each cluster. This capability enables us to compute a distance measurement for a source detection without a label relative to each class group in the cluster, based on the shapes of their distributions. This measurement is performed using the Equation \ref{eq:methods_eq6}.

We chose to use the \textit{Mahalanobis} distance as our metric for analysis because it takes into consideration the distributions of each class group within each cluster. This feature allows us to calculate the distance of a source detection without a label with respect to each class group in the cluster based on the shapes of their distributions. The computation is carried out using Equation \ref{eq:methods_eq6}.

After measuring the distance of a point $x$ to all classes in a cluster space, we apply the \texttt{softmin} function to convert the distances to a probability distribution. The \texttt{softmin} is a variation of the well known \texttt{softmax} function \citep{Goodfellow-et-al-2016}, and it is defined as:

\begin{equation}
    \label{eq:methods_eq10}
     \texttt{softmin}(x_n) = \frac{\exp(-x_n)}{\sum_j^N \exp(-x_j)}.
\end{equation}

where $x_n$ is a real number in a vector $\bm{x} = \{x_1, \dots, x_N\}$. $\texttt{softmin}(x_n)$ is equivalent to $\texttt{softmax}(-x_n)$. After applying this function to a vector, every element will be in the range $[0,1]$ and $\sum_n^N\texttt{softmin}(x_n)=1$. We apply the \texttt{softmin} function for all unlabeled points in all clusters, resulting in smaller distances yielding larger probabilities.  The final outcome is a table of source detections with a distribution of probabilities over classes. 

It is worth highlighting that a single source detection may receive multiple probabilistic classifications. This process can be represented mathematically as follows. Suppose we have $N$ detections for a source, denoted by $d_1, d_2, \ldots, d_N$. For each detection $d_i$, let $C_i$ be the set of possible classifications that it can receive. Then, we can define a collection of sets $C_1, C_2, \ldots, C_N$, where each set $C_i$ corresponds to the possible classifications for detection $d_i$. Thus, the relation is summarized as
$$d_i \rightarrow P(C_i)$$

\noindent where $P(C_i)$ represents the probability that detection $d_i$ belongs to each possible classification in the set $C_i$. This formalism represents classifications at the per-detection level, treating each detection as an individual target to classify.  We refine this process to produce a single classification for each source.

%$$\mathbf{x}^{source}_d \rightarrow %P_{classes}$$
%\noindent where $\mathbf{x}^{source}_d$ is a detection of a source, and is $P_{classes}$ is the set of probabilities for each considered class. We refine this process to produce a single classification for each source.

\subsubsection{Source master class}
\label{sec:master_class}

%In order to get a unique classification for each source in our dataset, we need to perform a \textit{master} classification that summarizes the different classifications of each detection of a source.

A fraction of the sources in the CSC have been observed more than once during the lifetime of \emph{Chandra}. A source with multiple detections will appear more than once in the catalog. The classification scheme that we have described so far operates on individual detections, which implies that sources with multiple observations get multiple classifications. While you expect classifications of the same source to agree, this is not always the case, since detections can differ in their properties. We therefore require a procedure to provide \emph{master} classification of sources, that takes into account the information from the multiple detection classifications.

To obtain the master class for a given source, we use two different approaches: \textit{soft} voting and \textit{hard} voting classifiers. In the \textit{hard} voting approach, also known as majority voting, we simply choose the most common classification among the different detections of a source. In the \textit{soft} voting approach, we profit from the probabilistic nature of the classification: for each detection of a source, we weight each class according to their probability, and then average over all detections in order to get the class with the highest mean probability. Both soft and hard voting have been widely used in ensemble methods of machine learning \citep{zhou2012ensemble}.

%Our master classification method gives us two important results: an \textit{agreeing} source classification table with sources that have the same classification in both the hard and soft voting methods, and a \textit{confused} source classification table with sources that have different classifications in both methods. 

We then arrange the results in two tables: a table of \emph{uniquely classified} sources, and a table of \textit{ambiguous} sources that yield different classifications with the two different methods.

To formalize the process of obtaining the master class for a given source, we introduce the following mathematical notation.

Let $S$ be a source, and let $\mathcal{D}_S$ be the set of all detections associated with $S$. The \textit{hard} voting classifier assigns the class $\mathcal{C}_S$ that appears most frequently among the detections in $\mathcal{D}_S$. That is,
$$\mathcal{C}_S = \text{argmax}_{c \in C} \sum_{d \in \mathcal{D}_S} [d \rightarrow c],
$$
where $C$ is the set of all possible classifications, and $[d \rightarrow c]$ is an indicator function that equals $1$ if detection $d$ was classified as $c$, and $0$ otherwise.

We have that $C_i$ is the set of all possible classifications for detection $d_i \in \mathcal{D}_S$. For each detection $d_i$, let $P_c(C_{i})$ be the probability that $d_i$ belongs to class $c$ in $C_i$.  We can then compute the \textit{soft} voting approach. We calculate the  mean probability for each class $c$ in $C$ across all detections, i.e., 
 
 $$C_S = \text{argmax}_{c \in C} \frac{1}{N_i} \sum_{i=1}^{N_i} P_c(C_{i}),$$ 

where $N_i = |C_i|$.

%For the \textit{soft} voting approach, we calculate a weight $w_{d,c}$ for each detection $d$ and class $c$, based on the probability that detection $d$ belongs to class $c$. Let $P(d)$ be the vector of probabilities for detection $d$ to belong to each possible classification, as defined earlier. Then, we can define the weight as $w_{d,c} = P(d)_c$. We can then compute the \textit{soft} voting classification for a source $s$ as
%$$c_s = \text{argmax}_{c \in C} \frac{1}{|\mathcal{D}_s|} \sum_{d \in \mathcal{D}_s} w_{d,c},$$
%where $|\mathcal{D}_s|$ is the number of detections for source $s$.

The full classification procedure that we have described here will be applied to the pre-processed CSC dataset, and their properties. Results of this process are presented in Section \ref{section:results}. A summary of our full pipeline is presented in Algorithm \ref{alg:classalg}.

%Results of this method are described in detail in Section \ref{section:results}, including an explanation of the reasoning behind the selection of this classification approach. 

\begin{algorithm}
$$
\textbf{A. Gaussian Mixture Model and Class Extraction}
$$

\noindent\textbf{1.} \textbf{\textit{Clustering} step:} Assign a cluster to all data points using a Gaussian Mixture Model.

\textbf{2.} \textbf{\textit{Crossmatch} step:} Crossmatch data with other databases (in this case, SIMBAD) to extract classes for all data points, if available.

$$
\textbf{B. Cluster-wise classification algorithm}
$$

% \noindent\textbf{1.} \textbf{\textit{Distance} step:} For a data point (detection) $\bm{x}$, compute the \textit{Mahalanobis distance} to all classes in its cluster. Store them as
% $$
% m_k(\bm{x}) = \{D_l(x) : \forall \;l \text{ in cluster }k\}.
% $$

% where $l$ represents a class label in the set of classes in a cluster.
\noindent\textbf{1.} \textbf{\textit{Distance} step:} For a data point (detection) $d_i$, compute the \textit{Mahalanobis distance} to all possible classes. Store them as
$$
m(d_i) = \{D(d_i) : \forall \;c \in C_i\},
$$

where $C_i$ represents all possible classes for detection $d_i$.

\textbf{2.} \textbf{\textit{Probability} step:} Convert the distances to a probability distribution over classes with

$$
\texttt{softmin}(m(d_i))
$$

\textbf{3.} Repeat for all registers in the data set.

$$
\textbf{C. \textit{Master} classification}
$$

\textbf{1.} \textbf{\textit{Hard} voting:} For each source $S$ in the data set, consider each of its detections and assign a class using the \textit{hard} voting approach, which selects the most common classification among the detections:
$$\mathcal{C}_S = \text{argmax}_{c \in C} \sum_{d \in \mathcal{D}_S} [d \rightarrow c],
$$

\textbf{2.} \textbf{\textit{Soft} voting:} For each source $S$ in the data set, consider each of its detections and assign a class using the \textit{soft} voting approach, which takes into account the probabilities of each detection belonging to each class:
$$C_S = \text{argmax}_{c \in C} \frac{1}{N_i} \sum_{i=1}^{N_i} P(C_{i,c}),$$ 

\textbf{3.} \textbf{Tables:} Aggregate the sources that agreed on their classification in both the \textit{hard} and \textit{soft} voting methods into a table of \textit{uniquely classified} sources. Aggregate the sources that were classified differently in each method in a table of \textit{ambiguous} sources.

\caption{Our proposed classification method. This technique takes advantage of a previous clustering output and use it to bound the classification to particular cluster spaces.}
\label{alg:classalg}
\end{algorithm}

\subsection{Hyperparameter tuning}
\label{sub:hyperparameter}
The most important hyper-parameter to determine is the number of components $K$ (\texttt{n\_components} in the \texttt{GaussianMixture} \textit{scikit-learn} architecture). In order to select the number of components, we preliminarily use the Bayesian Information Criterion (\textit{BIC}).

The \textit{BIC} was introduced in \cite{schwarzetal} and is defined as:

\begin{equation}
    \label{eq:eq4}
    \text{\textit{BIC} } = -2\log(\hat{L}) + \log(N)d
\end{equation}

where $\hat{L}$ is the maximum likelihood of the model, $d$ are the degrees of freedom or number of parameters estimated by the model, and $N$ is the number of samples or data points. The BIC measures how much variance in the data is explained by the incorporation of additional components. Additional components can help increase the likelihood, but beyond a certain complexity no additional information is gained, as we are overfitting. The BIC balances the goodness of fit of the model with the number of parameters. As the number of components in a GMM increases, the fit to the data improves, but the number of parameters also increases, leading to a more complex model. The optimal number of components is the one that provides a good balance between the fit to the data and the model complexity.

Usually a lower BIC is preferred. However, BIC decreases as the number of components increases, suggesting that the model is fitting the data better, but it also means that the model is becoming more complex and may be overfitting. We need to  choose the number of components with the lowest BIC score with a reasonable number of parameters. We explore this selection using the \textit{elbow method}.

The \textit{elbow method} is a well-known heuristic technique for determining the optimal number of components in a model. This method involves plotting the BIC (or originally, the dispersion) as a function of the number of components, and selecting the number of components where the decrease in BIC slows down, i.e., the number of components for which the slope of the tangent changes drastically. This indicates that adding additional components is not significantly improving the fit to the data. The elbow method was first introduced in \cite{thorndike1953belongs}. Recent studies have shown that the BIC-based component selection performs better than the traditional dispersion-based elbow method \citep{elbowno}.

\begin{figure}
\includegraphics[width=\columnwidth]{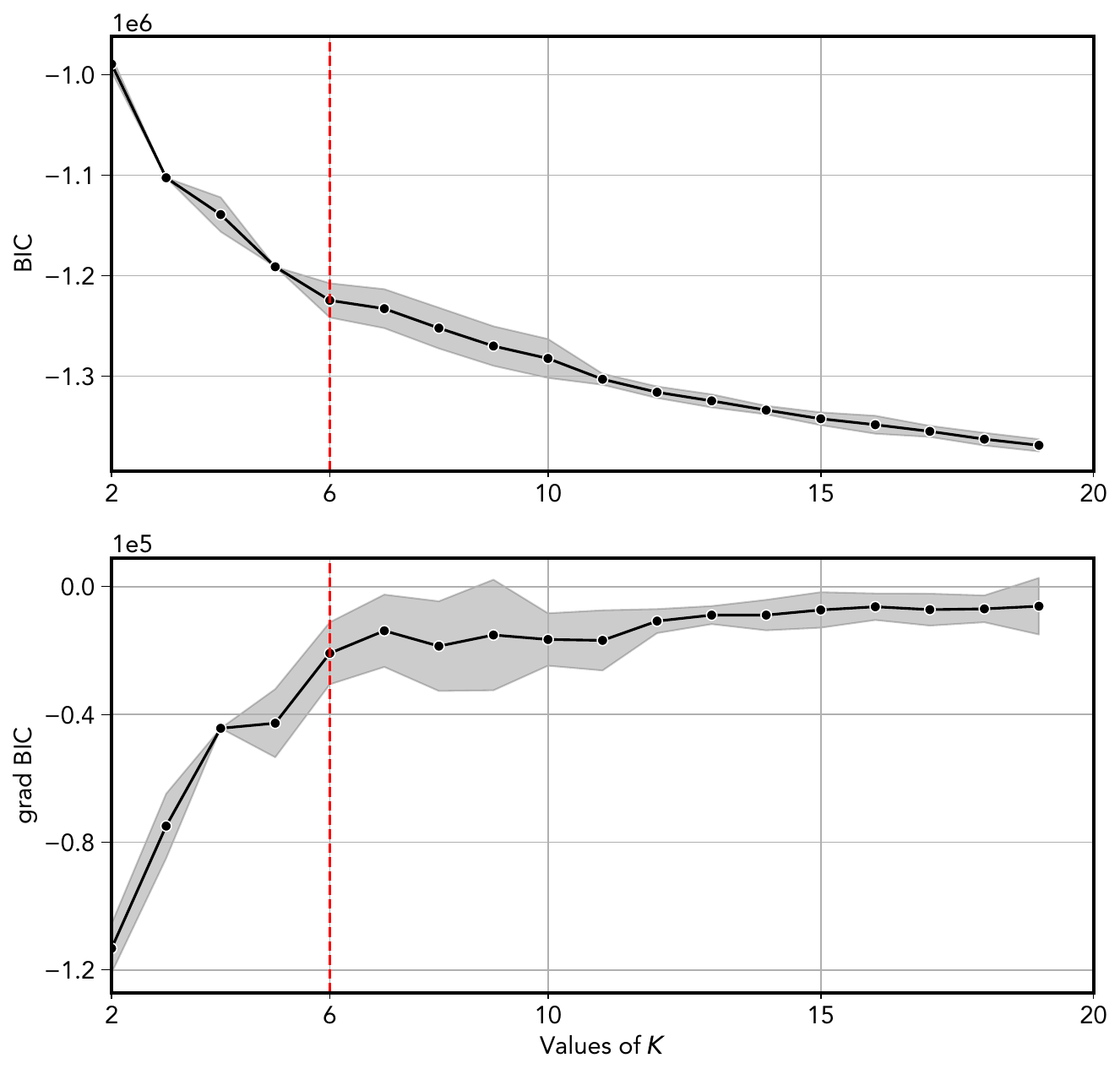}
\caption{The Bayesian Information Criterion (\textit{BIC}) (top) and the gradient of the \textit{BIC} (bottom) as a function of the number of components $K$. The number of components ranges from $2$ to $20$. The gray region delineates the confidence interval as determined by the standard deviation of each $K$'s iteration results. The \textit{BIC} function is smoothly decreasing, while its gradient shows a constant behaviour for values greater than $K=6$. The red dashed vertical line highlights the function point at $K=6$, which is the number of components that shows a better configuration in this technique.}
\label{fig:methods_1}
\end{figure}

We plot the BIC as a function of the number of components to discern where further addition does not substantially improve data fitting. Figure \ref{fig:methods_1} shows that the BIC plot lacks a clear elbow; therefore, we employ the gradient function to determine the optimal number of components. At $K=6$, we note a considerable change in the gradient's slope, which subsequently converges to 0 for $K \geq 7$. This observation indicates that augmenting the components beyond $K=6$ does not notably enhance the data fit.

In our analysis, we have used the BIC-based elbow method as an initial step guide. We have also relied on domain knowledge and a graphical analysis of the data distribution to confirm the number of components. We performed this analysis by experimenting with several number of components over the same dataset. This multi-faceted approach allows us to make informed decisions about the number of clusters and to ensure the reliability of our results. We provide an analysis on the stability of this selection, and its impact on the final classification in online Appendix C.

As a result of this process, we selected the number of components to be 6. Further graphical and distributional analysis revealed that the formation of the clusters was mainly influenced by a subset of the selected features (Table \ref{tab:properties}), specifically variability probabilities and hardness ratios. These findings are discussed in detail in Section \ref{section:results}.

Another crucial hyperparameter is the type of covariance parameters to use. These are defined by the \texttt{covariance\_type} parameter in the \texttt{GaussianMixture} method of \textit{scikit-learn}. It is challenging to determine the position and shape features of the different components a priori, i.e., we do not know how the covariance matrices of each Gaussian need to be in order to optimally fit the data points. Therefore, we chose the \textit{'full'} configuration, which allows each component to define its own covariance matrix (position and shape) independently \citep{scikit-learn}. To ensure reproducibility, we fixed the \texttt{random\_seed} of our Gaussian Mixture Model to $42$\footnote{"The answer to the ultimate question of life, the universe, and everything is $42$." \citep{adams1995hitchhikers}}.

\subsection{Validation}
Validating unsupervised learning techniques requires a departure from typical methods predominantly used in supervised learning research. The validation of our proposed algorithm not only confirmed its suitability for our classification problem but also guided us in assigning a 'master class' to unique sources. We do this by using a $70\%/30\%$ stratified random split over the set of SIMBAD labeled CSC detections. We treat the $30\%$ split as a benchmark set. These detections are treated as if they were unlabeled during the classification process, but we then compare the model's predictions to the known SIMBAD labels to assess its performance. This operation is performed using the \texttt{train\_test\_split} function from the \textit{scikit-learn} library. Its effectiveness is evaluated through the analysis of the resulting confusion matrices, which contrast the ground truth classes (SIMBAD) with our predicted classifications. These results are presented in \S~\ref{subsec:detectionlevel}.

%%% RESULTS
\section{Results}
\label{section:results}

In this section we present the results of applying the proposed pipeline to the CSC dataset described in \S~\ref{sec:data}. Using the GMM results, as well as the class associations using the Mahalanobis distance within each cluster, we assign probabilistic classes for each object detection in the dataset. We summarize classes for each individual object using hard and soft voting classification. We provide final \textit{uniquely classified} and \textit{ambiguous} classification tables for individual sources. We validate our results by constructing confusion matrices for a set of control detections, and by comparing our results to compiled catalogs of X-ray binaries, AGNs, etc.  We also quantify the uncertainty in our probabilistic classifications by evaluating the shape of the probability distribution over classes, and by evaluating consistency in the assigned class for bona-fide CSC sources and regions. We provide the code, data and results in the GitHub repository \href{https://github.com/samuelperezdi/umlcaxs}{https://github.com/samuelperezdi/umlcaxs}.

\subsection{Clustering}

We apply the Gaussian Mixture Model with the hyperparameters described in Section \ref{sub:hyperparameter} to the selected CSC features (\S~\ref{sec:data}) in order to identify an initial set of clusters. Members of a given cluster have relatively similar features among them, such as hardness ratios and variabilities. Features tend to be distinct from cluster to cluster, but this does not exclude significant overlap, or implies that all points within a cluster belong to the same class. We applied the clustering to a total of $29655$ individual CSC source detections. %Hereinafter, we examine how those detections are clustered and how we can use the topology of the objects within the clusters to assign probabilistic classes to those individual detections. Later, in \S~\ref{sec:discussion}, we offer a possible astrophysical interpretation of our results. Our results can be reproduced using the notebook \texttt{clustering} in our GitHub repository.

\begin{figure*}
\includegraphics[width=2\columnwidth]{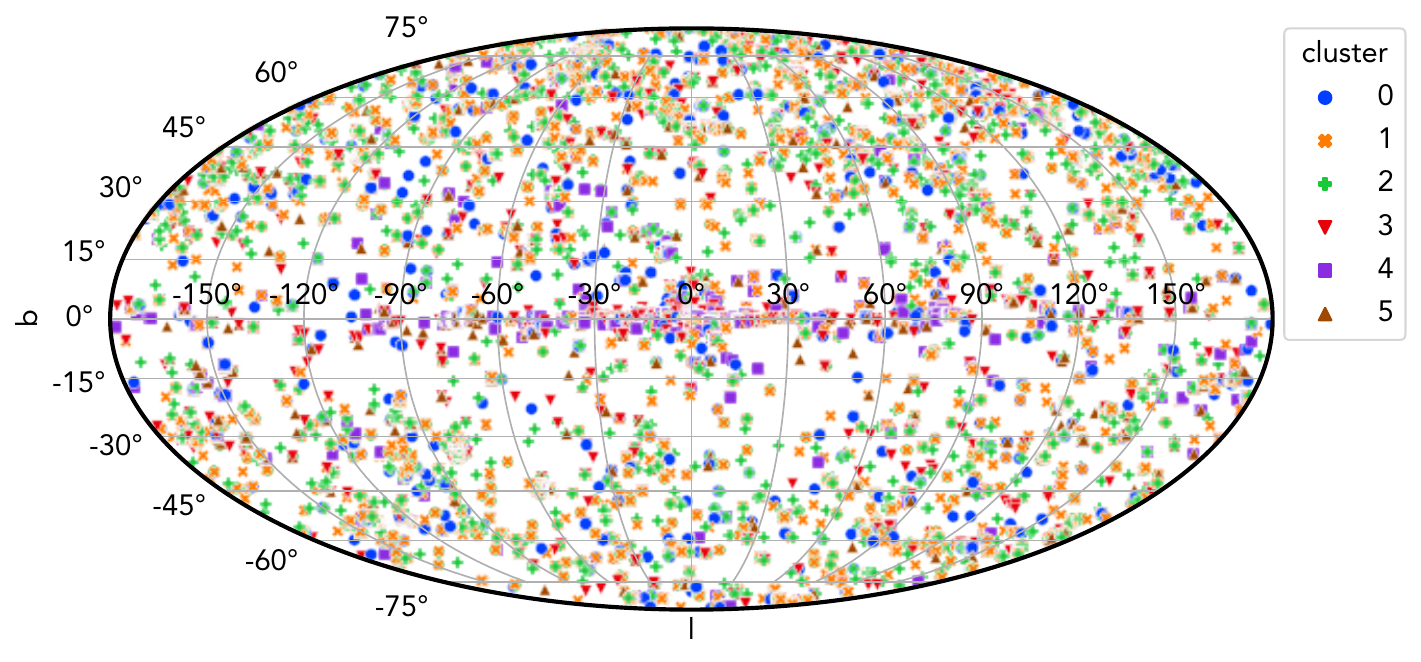}
\caption{Source detections in galactic coordinates over a Mollweide projection, discriminated by their assigned cluster in colors and markers. We see a trend of extragalactic or galactic for points in particular clusters.}
\label{fig:results_1}
\end{figure*}

%34.29
%28.78
%12.57
%9.76
Figure \ref{fig:results_1} shows a Mollweide projection of the sky, where the spatial distribution of the clustered detections is presented. We note a clear differentiation between sources located on the galactic plane, versus off-the-plane sources, as they tend to respectively be assigned to different clusters.  For instance, clusters 3 and 4 predominantly reside in the galactic plane, with $34.3\%$ and $28.8\%$ of detections at galactic latitudes $\lvert b \rvert < 5^\circ$, respectively. Conversely, clusters 2 and 1 predominantly consist of extragalactic source detections, with only $12.6\%$ and $9.8\%$ of detections at galactic latitudes $\lvert b \rvert < 5^\circ$, respectively. To a lesser extent, cluster 5 is also associated with off-plane sources, while cluster 0 exhibits a stronger connection to the galactic plane.

% Whether this is just an effect of more obscuration along the galactic plane, or a true difference in the populations of galactic versus extragalactic X-ray sources will be investigated later. 

\begin{figure*}
\includegraphics[width=2\columnwidth]{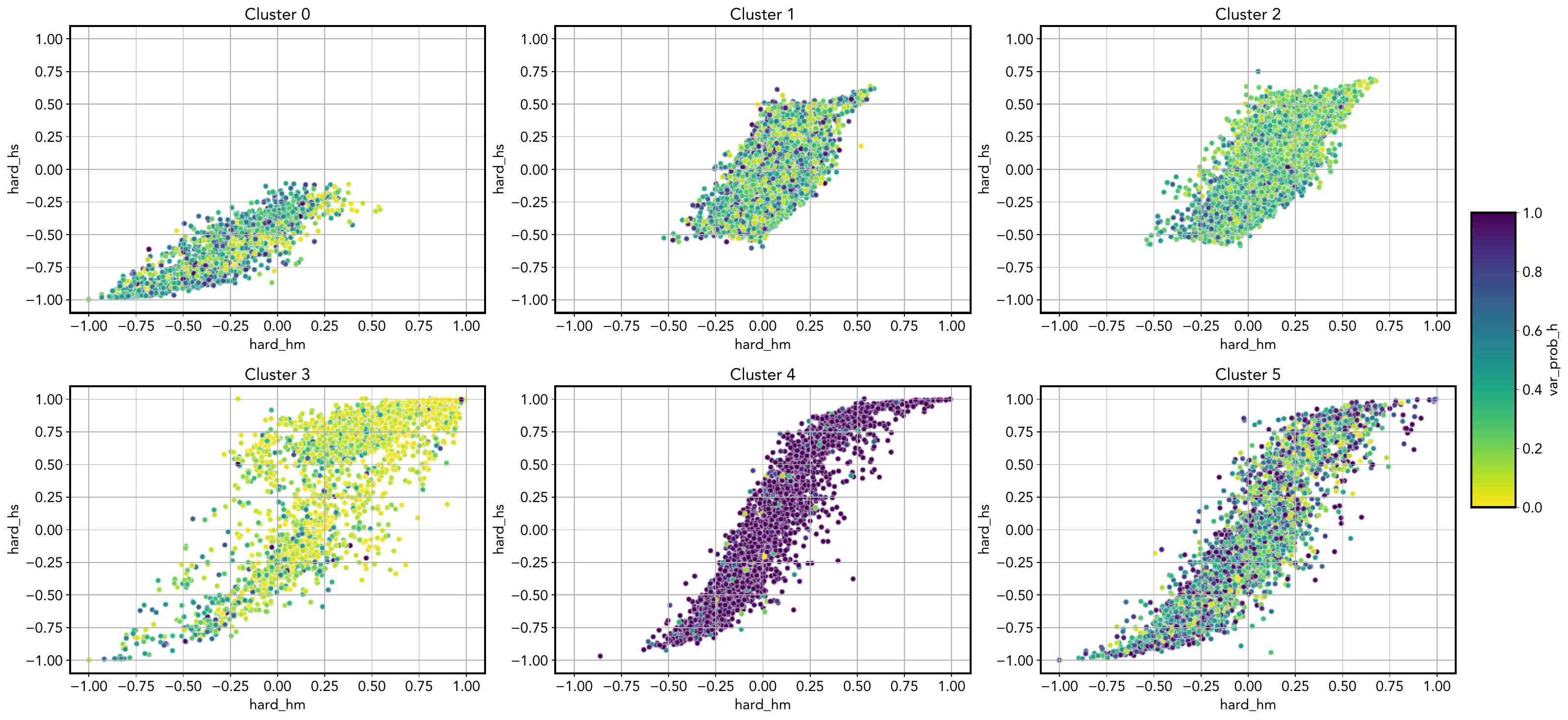}
\caption{\texttt{hard\_hs} vs. \texttt{hard\_hm} scatter plot for each cluster, with \texttt{var\_prob\_h} as a color dimension. Notice how some clusters clearly demonstrate multidimensional correlations, predominantly influenced by highly variable or low variable source detections.}
\label{fig:results_hardhshm_var}
\end{figure*}

In Figure \ref{fig:results_hardhshm_var} we show the hardness-to-hardness diagrams for objects in different clusters, color-coded by variability on the hard band. Overall, cluster membership relates to hardness and variability properties. Cluster 4 is made up almost entirely of highly variable objects irrespective of their spectral shape, whereas cluster 3 is predominately made of very low variability objects with a hard spectrum. Other clusters show more dispersion in variability, although with discernible differences in both the overall variability and the spectral shapes. Clusters 1 and 2 occupy an intermediate region in hardness. No single feature by itself can separate sources of different properties, but combined they can isolate specific behaviors.

%The feature-based cluster differentiation can also be seen in the histograms of Figures \ref{fig:hard_hs_hist} and \ref{fig:var_prob_hist}. 

%In the plots we have only explored a few projections of the multi-dimensional space of features. We note that the correlations between features that are relevant for the classification are not necessarily evident in the 2D projections that we have presented. This highlights the importance of employing a method that attempts the clustering in the multi-dimensional space of features, such as the GMM, in order to separate source detections based on the entire set of their properties.

%While our plots illustrate 2D projections of the multi-dimensional feature space, it is important to acknowledge that relevant classification correlations may not be evident in these visualizations. This underscores the significance of utilizing methods like GMM for clustering in the multi-dimensional feature space, enabling the separation of source detections based on their comprehensive property set.

% \begin{figure}
% \includegraphics[width=\columnwidth]{fig/results_4.pdf}
% \caption{\texttt{hard\_hs} vs. \texttt{hard\_hm} scatter plot, differentiated by cluster. This is a particular 2D projection of the multidimensional property space of our data. We can see some clear differences in the clusters, concentrating in distinct locations of the correlated distribution.}
% \label{fig:results_hardhshm}
% \end{figure}

%We also examine the distributions of specific properties to identify separations. Figures \ref{fig:hard_hs_hist} and \ref{fig:var_prob_hist} display histograms of the hardness ratios and variability properties (\texttt{hard\_hs} and \texttt{var\_prob\_b}). The separations previously observed are still evident. For instance, clusters 0 and 3 are differentiated into softer and harder detections, while clusters 1 and 2 concentrate around $\texttt{hard\_hs} = 0$. Cluster 4 is dominated by highly variable sources and stands apart from clusters 0, 1, 2, and 3. These low-dimension visualizations assist us in determining the properties crucial for cluster separation and verify previously observed separations. 

%\begin{figure}
%\includegraphics[width=\columnwidth]{fig/results_6a.pdf}
%\caption{Histogram plot for \texttt{hard\_hs}, normalized by bar height and colored for each cluster.}
%\label{fig:hard_hs_hist}
%\end{figure}

%\begin{figure}
%\includegraphics[width=\columnwidth]{fig/results_6b.pdf}
%\caption{Histogram plot for \texttt{var\_prob\_b}, normalized by bar height and colored for each cluster.}
%\label{fig:var_prob_hist}
%\end{figure}

\subsection{Classification}

So far, our clusters contain a mix of SIMBAD-classified objects and a majority of unclassified objects. We first investigate if there are dominant classes within each cluster In \autoref{tab:cpred} we rank the 10 most common classes within each cluster. Unclassified detections are marked as \textit{NaN}, and are the majority of data points in all clusters, except for cluster 4. Some of the clusters are clearly dominated by object detections of similar astrophysical types. Such is the case of cluster 4, for which over 90\% of the classified objects are of types associated with young stars. Clusters 0 and 5 also appear to be dominated by young stars, whereas for cluster 2 practically the entire set of SIMBAD-classified objects corresponds to either super-massive black holes (QSO, AGN, Seyfert), or stellar size black holes (HMXB, XB). The X-ray properties of accreting compact systems are expected to be similar, but we find clusters (e.g. cluster 1) where AGNs and QSOs are significantly more common than X-ray binaries. Despite these trends, clusters are not generally populated by objects of a given class. %The X-ray properties of accreeting compact systems are expected to be similar, although we will later investigate to what extent AGN-type objects can be distinguished from X-ray binaries. For other clusters, there is a mix of types, such as cluster 1, that has comparable numbers of young stars and accreeting objects. We will later investigate under which conditions or evolutionary stages these different classes can have similar X-ray properties. As we can see, we use clustering first step on the classification algorithm. Clustering groups objects of similar properties, which facilitates the classification step.

\subsubsection{Classes}
\label{sec:classes}
Some of the classes represented in the SIMBAD database are extremely rare, having less than 10 examples each. Our initial BIC test, on the other hand, indicates that the X-ray properties alone contain only a limited amount of information to separate the objects, represented by the optimal number of GMM clusters selected (6). It would therefore be meaningless to attempt a classification using all the 125 classes that appear in the SIMBAD database for our sources. For training, we exclude sources in these underrepresented classes. The remaining 10 classes that we use for training, representing a broad range of astrophysical phenomena are: \textit{QSO}, \textit{AGN}, \textit{Seyfert\_1}, \textit{Seyfert\_2}, \textit{HMXB}, \textit{LMXB}, \textit{XB}, \textit{YSO}, \textit{TTau*}, \textit{Orion\_V*}. We can further separate these classes in 4 larger groups, which we will refer as aggregated classes:

\begin{itemize}
    \item \textit{\textbf{Large accretors}} \textbf{(AGN)}: \textit{QSO} and \textit{AGN}. 
    \item \textit{\textbf{Extended Large accretors with optical lines}} \textbf{(Seyfert)}: \textit{Seyfert\_1} and \textit{Seyfert\_2}
    \item \textit{\textbf{Small accretors}} \textbf{(XB)}: \textit{HMXB}, \textit{LMXB}, and \textit{XB}
    \item \textit{\textbf{Young stars}} \textbf{(YSO)}: \textit{YSO}, \textit{TTau*}, and \textit{Orion\_V*}
\end{itemize}

In the context of this study, "accretors" refer to accreting compact objects. In "Large accretors", the central object is a super-massive black hole, which results in a high luminosity. "Small accretors" involve a stellar remnant (a neutron star or black hole) accreting matter from a companion star.

For a significant number of X-ray sources, SIMBAD only provides a general class that is not informative as of specific physical properties (\textit{"Star"}, \textit{"X"}, \textit{"Radio"}, \textit{"IR"}, \textit{"Blue"}, \textit{"UV"}, \textit{"gamma"}, \textit{"PartofG"}). We assume objects in those classes to be unclassified, and assign them classes using our algorithm. 

%So far, we have selected the output classes of our pipeline. Given the big diversity in the SIMBAD matching classes, we decided to select source detections with matching \textit{ambiguous} types to be included in the target dataset, i.e., the source detections that were classified by our pipeline. These classes were selected taking into account the lack of information beyond the peculiar emission of the source, errors in the classification, or very broad types that do not include information about the nature of the source. The final classes that were considered as \textit{ambiguous} are \textit{Star}, \textit{X}, \textit{Radio}, \textit{IR}, \textit{Blue}, \textit{UV}, \textit{gamma}, \textit{PartofG}, \textit{**}. In other words, source detections that had one of these matching classes, were considered a target to classify. Selection of these classes was made extensively considering the SIMBAD documentation, and making sure that the classes were present as source detections in our dataset. We will refer these classes often as \textit{ambiguous} classifications. 

All in all, the total number of CSC detections for which we will provide a label is $14,507$.

%The selection of classes also implies a reduction of the original \textit{C-CSC} dataset. Source detections with target classes, ambiguous classes or \textit{NaN} labels represent an $84\%$ of the \textit{C-CSC} dataset, which has in total $29655$ source detections. We will refer to this dataset as the classification dataset, abbreviated as \textit{CSCclass}. 

%\textit{NaN} labels and ambiguous source detections represent the $58\%$ of the \textit{CSCclass} dataset, which is a number of $14507$ source detections. These are the source detections that we classified and present in the final output. Source detections in the \textit{CSCclass} dataset represent a total of $8756$ unique sources. 

\begin{table*}
\begin{multicols}{3}
    \begin{subtable}[h]{0.2\textwidth}
    \centering
        \begin{tabular}{lr}
        \toprule
        main\_type & size  \\
        \midrule
        NaN       &   884 \\
        Star      &   310 \\
        Orion\_V*  &   310 \\
        YSO       &   264 \\
        QSO       &   185 \\
        TTau*     &   165 \\
        X         &   144 \\
        PartofG   &   105 \\
        Seyfert\_1 &    52 \\
        SB*       &    47 \\
        \bottomrule
        \end{tabular}
        \caption{Cluster 0.}
    \end{subtable}\par%

    \begin{subtable}[h]{0.2\textwidth}
    \centering
        \begin{tabular}{lr}
        \toprule
        main\_type &    size   \\
        \midrule
        NaN       &  2869 \\
        QSO       &   931 \\
        X         &   887 \\
        Orion\_V*  &   304 \\
        YSO       &   304 \\
        Star      &   277 \\
        AGN       &   236 \\
        HMXB      &   210 \\
        Seyfert\_1 &   172 \\
        LMXB      &   141 \\
        \bottomrule
        \end{tabular}
        \caption{Cluster 1.}
    \end{subtable}\par

    \begin{subtable}[h]{0.2\textwidth}
    \centering
        \begin{tabular}{lr}
        \toprule
        main\_type     &   size    \\
        \midrule
        NaN           &  3211 \\
        X             &  1093 \\
        QSO           &  1081 \\
        Star          &   398 \\
        HMXB          &   272 \\
        GlCl          &   255 \\
        Seyfert\_1     &   252 \\
        XB*\_Candidate &   238 \\
        AGN           &   232 \\
        XB            &   222 \\
        \bottomrule
        \end{tabular}

        \caption{Cluster 2.}
    \end{subtable} \par  
\end{multicols}

\begin{multicols}{3}

    \begin{subtable}[h]{0.2\textwidth}
    \centering
        \begin{tabular}{lr}
        \toprule
        main\_type &   size    \\
        \midrule
        NaN       &   821 \\
        X         &   307 \\
        HMXB      &   165 \\
        YSO       &   154 \\
        Star      &   135 \\
        Seyfert\_2 &   120 \\
        QSO       &   115 \\
        Orion\_V*  &    91 \\
        Seyfert\_1 &    79 \\
        Pulsar    &    73 \\
        \bottomrule
        \end{tabular}

        \caption{Cluster 3.}
    \end{subtable}\par

    \begin{subtable}[h]{0.2\textwidth}
    \centering
        \begin{tabular}{lr}
        \toprule
        main\_type     &    size   \\
        \midrule
        Orion\_V*      &   688 \\
        YSO           &   562 \\
        NaN           &   429 \\
        Star          &   312 \\
        TTau*         &   159 \\
        X             &    78 \\
        HMXB          &    63 \\
        YSO\_Candidate &    62 \\
        BYDra         &    62 \\
        Em*           &    26 \\
        \bottomrule
        \end{tabular}
        \caption{Cluster 4.}
    \end{subtable}\par

    \begin{subtable}[h]{0.2\textwidth}
    \centering
        \begin{tabular}{lr}
        \toprule
        main\_type &   size    \\
        \midrule
        NaN       &  1131 \\
        YSO       &   563 \\
        Orion\_V*  &   530 \\
        Star      &   332 \\
        X         &   313 \\
        TTau*     &   196 \\
        QSO       &   187 \\
        HMXB      &   161 \\
        AGN       &    86 \\
        Seyfert\_1 &    71 \\
        \bottomrule
        \end{tabular}
        \caption{Cluster 5.}
    \end{subtable}\par 

\end{multicols}

\caption{The 10 most prevalent source classes in each of the clusters, ranked by number of examples. Source detections with non matching types are labeled as \textit{NaN}.}
\label{tab:cpred}
\end{table*}

\subsubsection{Detection level classification}
\label{subsec:detectionlevel}

%We obtain probabilistic classifications at the detection level. That is, for each CSC source that has been observed multiple times, we obtain one posterior probability over classes for each of the observations of that source. From the $8756$ unique sources in the data, $6802$ ($77.7\%$) have 1 detection, $1000$ ($11.4\%$) have 2 detections, $882$ ($10.1\%$) have between $3$ and $10$ detections, and finally just $72$ ($0.8\%$) sources have more than $10$ detections. For variable sources, and in particular for spectrally variable sources, the class that maximizes the posterior is not necessarily the same for all observations. For example, X-ray binaries caught in different states of their accretion cycles can look spectrally very different in the different states. It is therefore important to assess how specific properties (variabilities, hardness ratios, etc.) trace to specific classes at the observation level. We expect that regardless of the spectral changes over time, the distribution of posterior classes over time for a given source will be distinctive enough to assign reliable class. We described how we assign a final master class to sources that have been observed multiple times in Section \ref{sec:methods}, and we will discuss the results of that process later. We now focus on how the specific properties relate to the different classes.

Using the probabilistic approach described in \S~\ref{sec:methods_2} we have assigned a probabilistic class to all 14,507 CSC detections. 

\begin{figure*}
\vskip 0.2in
\begin{center}
\centerline{\includegraphics[width=2\columnwidth]{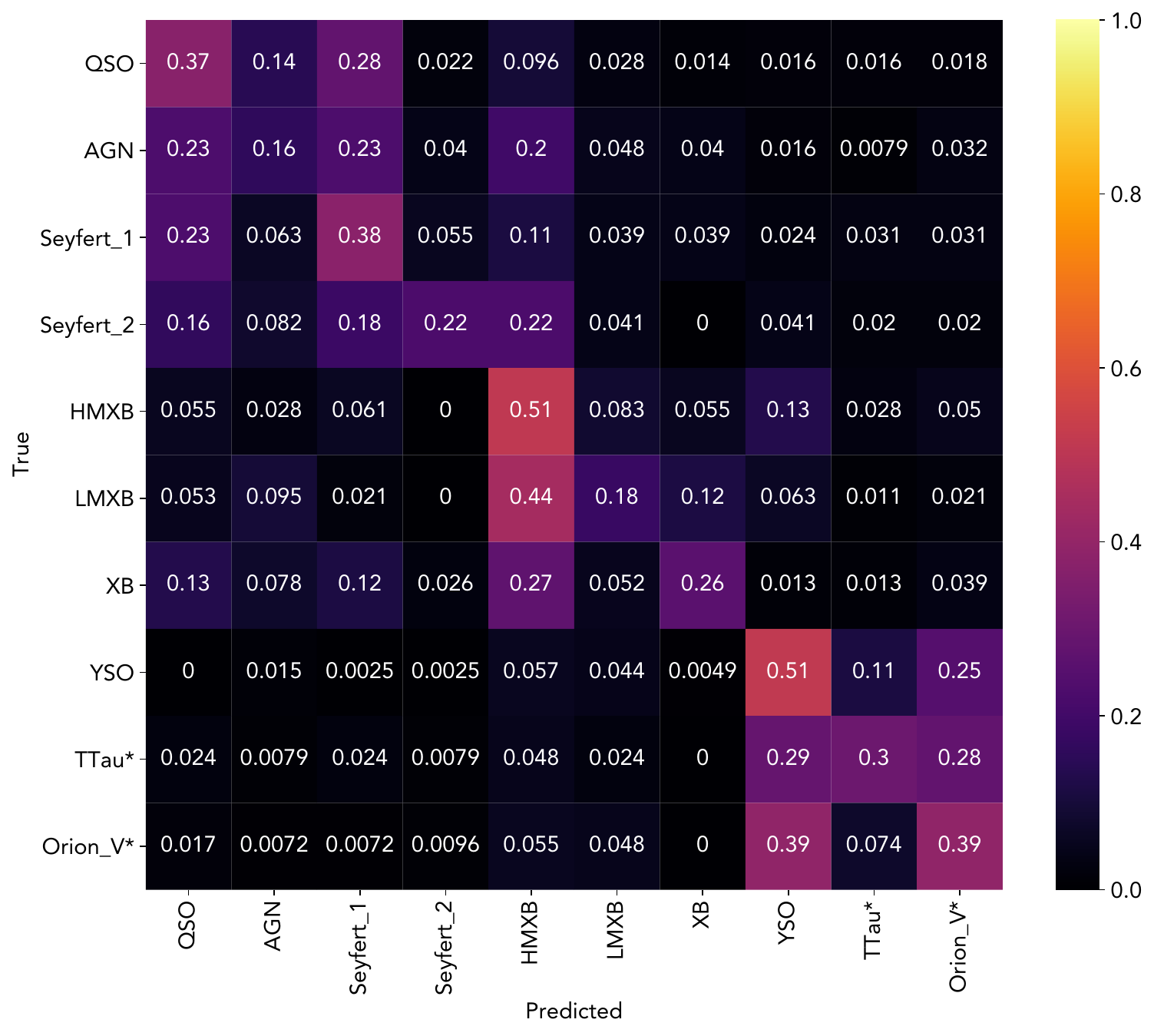}}
\caption{True vs. Predicted confusion matrix for the benchmark set using individual detections only, normalized by row. For each true class the proportion of source detections of that class assigned to each of the possible labels is shown.}
\label{fig:results_class_confusion_raw_out}
\end{center}
\vskip -0.2in
\end{figure*}

Figure \ref{fig:results_class_confusion_raw_out} shows the True vs. Predicted confusion matrix for the benchmark set previously described. Each element of the matrix indicates the fraction of test objects with a given True class that have been correctly classified by the algorithm. At the detection level, our X-ray based classification method is successful at distinguishing between large accretors (QSO, AGN, and Seyferts), small accretors (X-ray binaries), and young stars (YSOs, T-Tauri stars, and Orion V* variables). However, a fair amount of confusion persists between sub-classes of each major class.

\begin{figure}
\vskip 0.2in
\begin{center}
\centerline{\includegraphics[width=1\columnwidth]{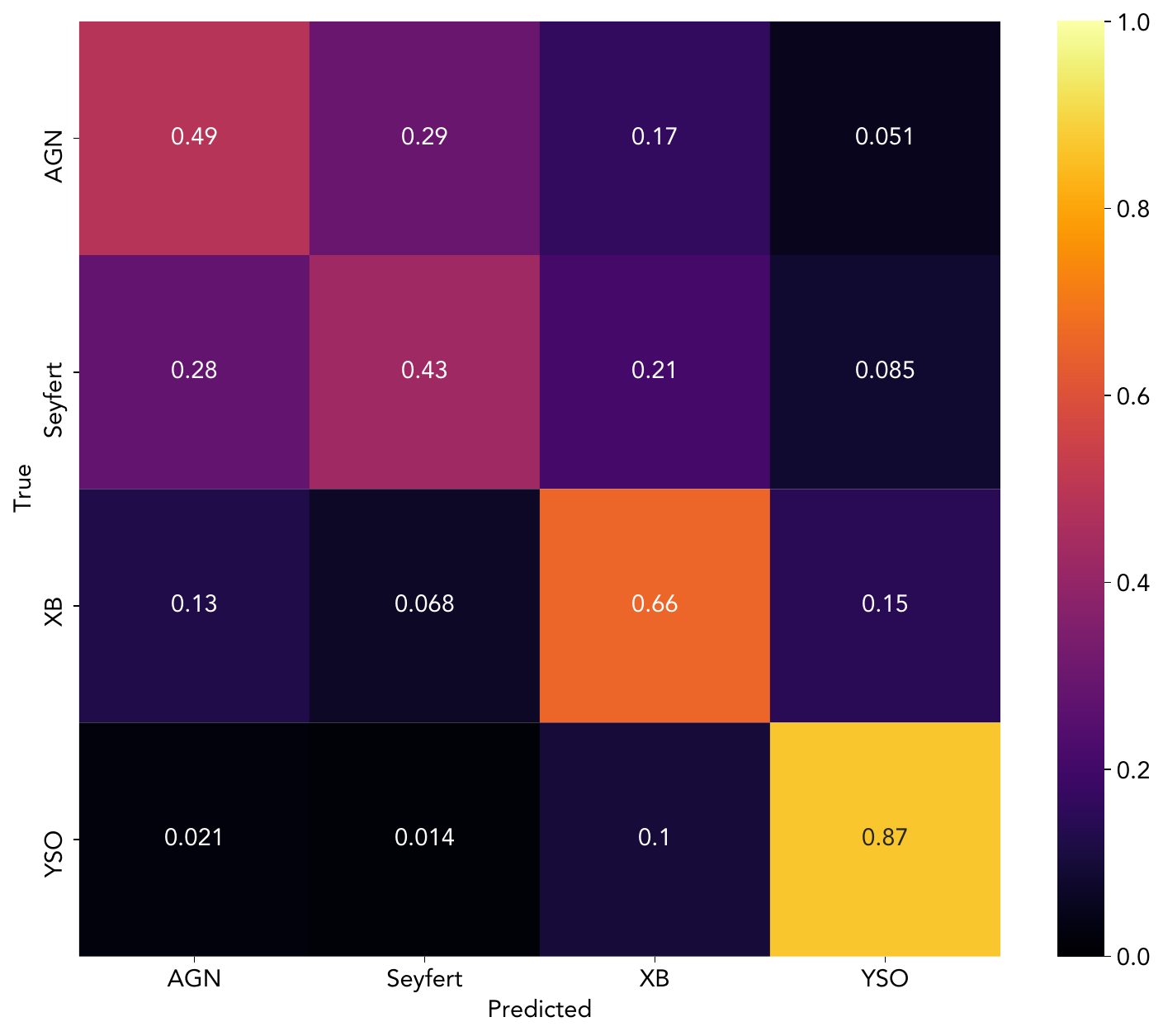}}
\caption{True vs. Predicted confusion matrix for the benchmark set, normalized by row. Classes were replaced by the aggregated classes \textbf{AGN}, \textbf{Seyfert}, \textbf{QSO}, \textbf{YSO}. The matrix displays the proportion of source detections in a specific class that were correctly classified or misclassified. The data used in this plot is not a final result of our pipeline.}
\label{fig:results_class_confusion_raw_grouped_out}
\end{center}
\vskip -0.2in
\end{figure}

For example, A QSO is correctly classified as a such about 40\% of the times, whereas about 30\% of the times, the algorithm thinks it is a Seyfert 1 galaxy. In fact, QSO X-ray spectra can be similar to Seyfert 1 spectra \citep{dadina2016xmm}. Lacking a distance or a redshift, the algorithm cannot distinguish between the two in the detection level. Similarly, a HMXB gets correctly classified over 50\% of the time, and incorrectly classified as a LMXB in about 8\% of the cases. On the other hand, LMXBs are classified as HMXBs in 44\% of the test examples. Among young stars, most confusion occurs for T-Tauri stars, which are classified as such only a third of the times, with the remaining examples classified in equal amounts as either YSOs or Orion V*.

Large accretors are most likely to be confused with HMXB, but that only happens about 20\% of the time, and small accretors are almost never assigned a large accretor label, with the exception perhaps of the XB class, which is confused with QSOs in 13\% of the examples. The most robust classification is provided for young stars, which are almost never classified in other groups. This is illustrated in Figure \ref{fig:results_class_confusion_raw_grouped_out}, where we have grouped individual classes into major classes 

A significant part of the confusion is likely due to differences in the spectral states of the different objects during different epochs. More robust classifications result from considering multiple detections of a source, but we note that single-epoch, X-ray-only classification of CSC sources is possible at some significance level.

%For each unclassified source detection, we obtain a distribution of probabilities over selected original classes. The result of this exercise is a probabilistic classification (i.e. a posterior over classes) for $14507$ CSC source detections from $8756$ unique sources. This means that a source with different detections can potentially have different classifications. We aimed for this in order to pinpoint exotic phenomena, and also to compare how the behaviour of a source changes for different detections, making it look like a member of different classes at different times. The classification table, named \texttt{classified\_cscs.csv}, and its respective code (\texttt{classification.ipynb}), is available at the provided \texttt{GitHub} repository.

%In this subsection, we are going to explore this first result, which corresponds to a classification for source detections. 

\begin{figure*}
\includegraphics[width=2\columnwidth]{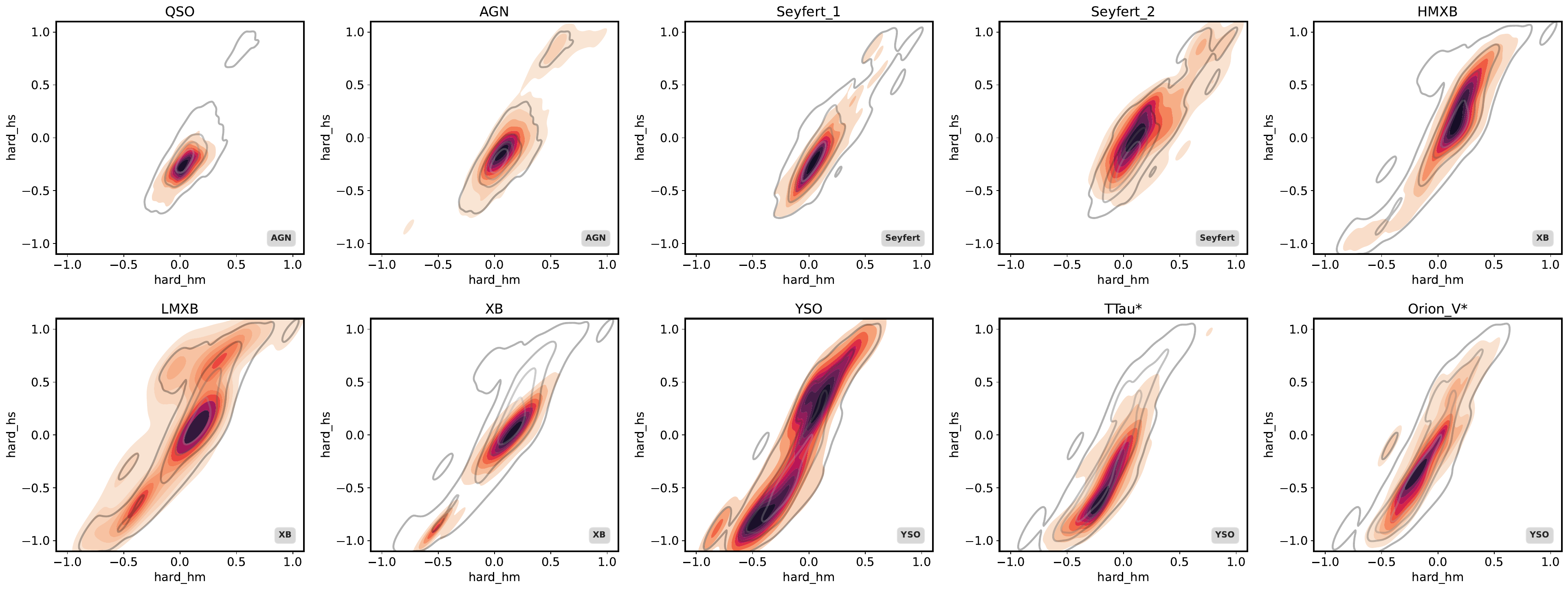}
\caption{\texttt{hard\_hs} vs. \texttt{hard\_hm} distribution density estimation for each class group, visualized using a kernel density estimate (KDE). Darker hues represent areas of higher concentration within the distribution. Each corresponding aggregated class density is included, delineated by gray contours and referenced in the gray boxes. Note that the density estimation may appear to exceed the valid value boundaries due to the smoothing effect. This effect does not indicate actual data points outside the hardness ratio limits but results from the bandwidth choice in the estimation.}
\label{fig:results_class_exp_hard}
\end{figure*}

How do X-ray properties distribute between classes? Figure \ref{fig:results_class_exp_hard} shows a density map of the distribution of hardness ratios for the classified detections. We observe relatively distinct distributions in the spectral shape for each of the different classes, even if there are also similarities and degeneracies. For example, QSOs, AGNs, and Seyfert 1 galaxies all appear to have a similar mean in their hardness ratio distributions, but they differ in the standard deviations of both hardness ratios. Seyfert 2 galaxies, on the other hand, have a slightly elevated hard-to-soft ratio with respect to the other three groups. X-ray binaries show wider distributions of their hardness ratios, with low-mass X-ray binaries even showing signs of multi-modality. This is relevant in the distinction between large and small accretors. Young stars show overall softer spectra with respect to compact objects, although in the case of YSO there seems to be at least two separate populations with different spectral shapes. T-Tauri stars have consistently lower hardness ratios, whereas Orion variables, which tend to show eruptive behavior, appear to occupy a region in between the two populations of YSO types.

\begin{figure*}
\includegraphics[width=2\columnwidth]{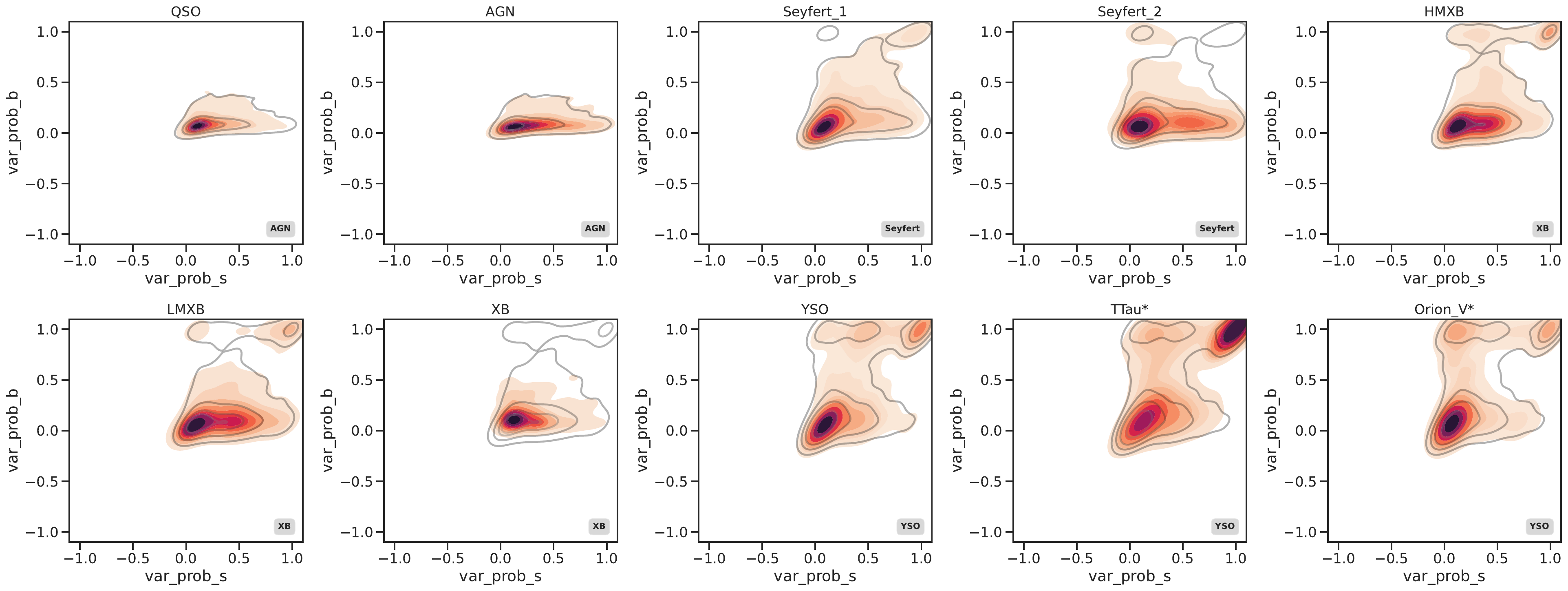}
\caption{\texttt{var\_prob\_b} vs. \texttt{var\_prob\_s} distribution density estimation for each class group, visualized using a kernel density estimate (KDE). Darker hues represent areas of higher concentration within the distribution. Each corresponding aggregated class density is included, delineated by gray contours and referenced in the gray boxes. Note that the density estimation may appear to exceed the valid value boundaries due to the smoothing effect.  This effect does not indicate actual data points outside the probability limits but results from the bandwidth choice in the estimation.}
\label{fig:results_class_exp_var}
\end{figure*}

Figure \ref{fig:results_class_exp_var} shows a similar density map, but this time for the variabilities in the broad (0.5keV-7.0keV) and soft (0.5keV-1.2keV) bands. We note significant variations in the level of soft-band variability across classes, with less variability in the integrated broad band. For most of the classes, variability is not correlated between bands, indicating spectral variability. Among large accretors, QSOs appear less variable in the soft band with respect to AGN, and Seyfert 2 galaxies appear more spectrally variable than their Seyfert 1 counterparts, which also show more variability in the broad band. For small accretors, no significant changes in variability is discernible between HMXBs and LMXBs, whereas sources classified plainly as XBs show a smaller degree of soft band variability. Finally, for young stars, T-Tauri stars are clearly differentiated by a marked bimodal distribution, and the majority of objects having a significant variability probability (\verb+var_prob_b ~ 1+). The latter might be related to flaring events in that stage of their evolution.

\subsubsection{Master classification}

So far we have obtained probabilistic classifications for every detection of a source in our dataset. In order to obtain more accurate classes for a given CSC source, we can now operate on the probabilistic labels assigned to each detection of a source. Spectral variability, different exposure times or signal to noise level might translate into different classifications in different epochs. From the $8756$ unique sources in the data, $6802$ ($77.7\%$) have only 1 detection (for these sources, the detection level classification is our best guess), $1000$ ($11.4\%$) have 2 detections, $882$ ($10.1\%$) have between $3$ and $10$ detections, and finally just $72$ ($0.8\%$) sources have more than $10$ detections. For each source we now assign a master class using the procedure described in \S~\ref{sec:master_class}.

Overall, both the hard and soft voting systems agree with each other: for a given source, the highest average probability across classes usually aligns with the primary type assigned to the majority of a source's detections. Only $485$ sources (6\% of the total) result in different classifications. The Cohen's Kappa coefficient \citep{cohen1960coefficient}, a measurement of agreement between the two methods, was calculated to be $0.94$. We compile a table of \textit{uniquely classified} and a table of \textit{ambiguous class} sources depending on whether the class assigned agree between the two methods. We also provide the mean classification probabilities and their standard deviations.  In the \textit{uniquely classified} sources table, \texttt{master\_class} refers to the class assigned from both the soft and hard classifier, and \texttt{agg\_master\_class}  denotes its corresponding aggregated class. In the \textit{ambiguous} sources table, \texttt{soft\_master\_class} and \texttt{hard\_master\_class}, refer respectively to each of the methods. We provide the number of detections per source as \texttt{detection\_count}. Sample extracts of the \textit{uniquely classified} and \textit{ambiguous} classification tables, sorted by detection count, are presented in Table \ref{A:table_confident} and Table \ref{A:table_confused} respectively. Full versions of the tables are available in the online supplementary material and the provided GitHub repository.

%Figure \ref{fig:results_galactic_classes} presents an overview of classified sources across galactic coordinates, leveraging a Mollweide projection to convey their spatial distribution across the sky.  We annotate regions of interest that have been previously investigated by the Chandra X-ray Observatory.

\subsubsection{Astrophysical validation of the classification}

\begin{figure*}
\vskip 0.2in
\begin{center}
\centerline{\includegraphics[width=2\columnwidth]{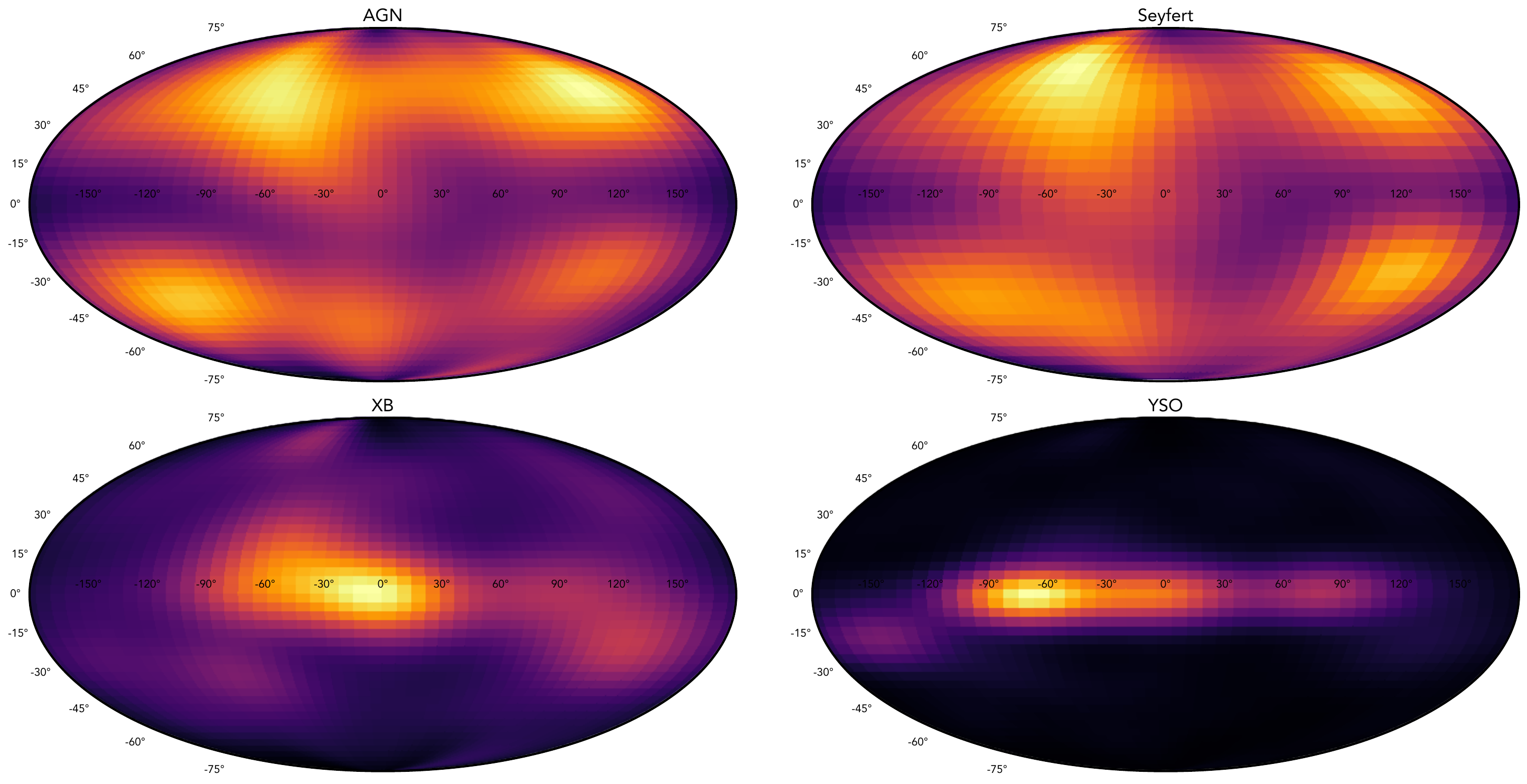}}
\caption{Density map for the \textit{uniquely classified} CSC sources that lacked a label before our study, in galactic coordinates, over Mollweide projections for each aggregated master class.}
\label{fig:results_galactic_classes_dens}
\end{center}
\vskip -0.2in
\end{figure*}

Certain types of CSC sources are expected to occupy specific regions of the sky. For example, we expect YSOs to be associated with star-forming regions along the galactic plane, and AGNs and quasars to be mostly located off the galactic plane. As a way to validate our classification, we investigate whether this expectations are satisfied -in a statistical sense- for the objects that we have classified. Figure \ref{fig:results_galactic_classes_dens} shows density maps for each aggregated class over Mollweide projections. We note that objects assigned to the \textbf{YSO} class clearly concentrate along the galactic plane, while AGNs are mostly confined to the extragalactic area of the sky. Objects that we have classified as X-ray binaries have a tendency to be located along the galactic plane, but not as starkly as in the case of YSOs. Seyfert galacies are mostly extragalactic, but less markedly than AGNs. These maps provide an early indication that our classifier is associating the right objects to the right environments, despite not having access to any information about the object's location in the sky. It is important to note that \emph{Chandra} is not an all-sky survey, and therefore these maps should not be interpreted as proxies for the actual surface densities of each class.

% \begin{figure*}
% \vskip 0.2in
% \begin{center}
% \centerline{\includegraphics[width=2\columnwidth]{fig/anno_results_galactic_classes.pdf}}
% \caption{\textit{Agreeing classification} sources, in galactic coordinates, over a Mollweide projection, color coded by aggregated master class.}
% \label{fig:results_galactic_classes}
% \end{center}
% \vskip -0.2in
% \end{figure*}

\begin{figure}
\vskip 0.2in
\begin{center}
\centerline{\includegraphics[width=1\columnwidth]{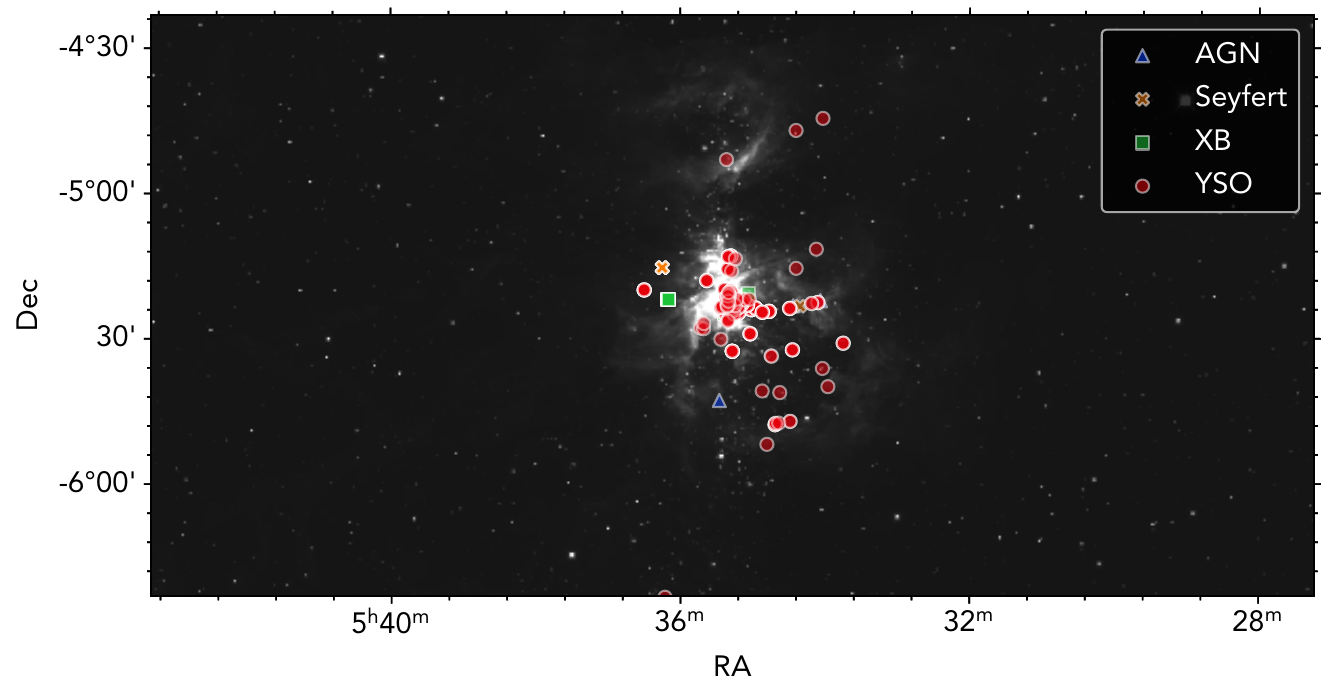}}
\caption{A set of previously unclassified sources in \textit{M42}, overlaid on an AllWISE W1 band image in grayscale. The colors indicate different classes assigned by our algorithm.}
\label{fig:m42}
\end{center}
\vskip -0.2in
\end{figure}

This association between assigned classes and environment holds even at smaller scales. Figure \ref{fig:m42} shows previously unclassified sources within the Orion Nebula region (M42) to which we have assigned a master class. Most of the sources in this area are classified as \textbf{YSO}, as expected in this well known star-forming region \citep{o2001orion}. We also observe a few sources classified differently. We now discuss two examples of these:

\begin{itemize}
    \item \textit{2CXO J053516.0-052353} has 4 CSC detections and is classified as an AGN. Figure \ref{fig:2CXO_J053516.0-052353} shows its probability distribution across aggregated classes. There is significant confusion between classes, with the AGN and YSO being equally probable. In observation ID $4374$, the source was classified as a \textit{YSO} with a high probability of $0.99$, and in observation ID $3744$, it was classified as \textit{Orion\_V*} with a probability of $0.75$.  In the remaining two observations, the source was classified as \textit{AGN} with probabilities around $0.7$, which resulted in the final hard-voting class. Furthermore, it is worth noting that the standard deviation is higher for the \textit{YSO} class ($0.41$) compared to the \textit{AGN} class ($0.38$). Such high standard deviations are not uncommon in our classification scheme. In summary, this is a close call between YSO and AGN class, which would still be classified as a Young Star if only the aggregated classes defined in \S~\ref{sec:classes} were used.
    
    \begin{figure}
    \vskip 0.2in
    \begin{center}
    \centerline{\includegraphics[width=1\columnwidth]{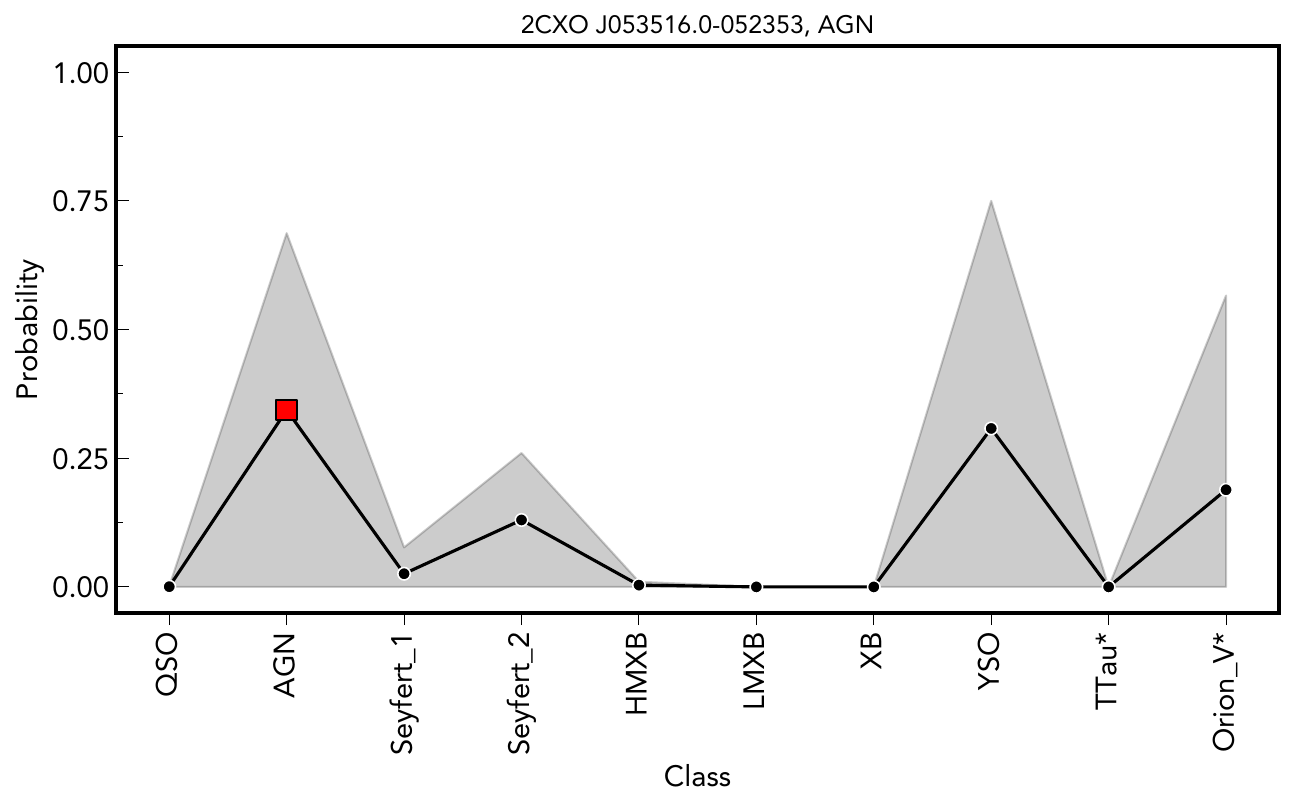}}
    \caption{Probability distribution over classes for \textit{2CXO J053516.0-052353}. The red square indicates the final master class. The gray area is the $95\%$ confidence interval and the black dotted line is the average probability when all detections are aggregated.}
    \label{fig:2CXO_J053516.0-052353}
    \end{center}
    \vskip -0.2in
    \end{figure}
    
    \item \textit{2CXO J053615.0-051530} has 5 CSC detections and is classified as Seyfert\_1 galaxy, with relatively high probabilities also assigned to the QSO and AGN classes. The closest match in the SIMBAD database, source \textit{V* V828 Ori}, is as an Orion Variable located about 3 arcsec to the southwest of the CSC source. In \cite{szegedi2013new}, the source is included in the \textit{TTau*} group, and other general variable and proper motion stars catalogs. We cannot rule out the possibility that the CSC detection might be in fact a background AGN.
    
    %     \begin{figure}
    % \vskip 0.2in
    % \begin{center}
    % \centerline{\includegraphics[width=1\columnwidth]{fig/2CXO J053615.0-051530.pdf}}
    % \caption{Distribution of probabilities over classes for the source \textit{2CXO J053615.0-051530}. The red square indicates the final master class. The gray area is the $95\%$ confidence interval and the black dotted line is the average probability aggregating all detections.}
    % \label{fig:2CXO_J053615.0-051530}
    % \end{center}
    % \vskip -0.2in
    % \end{figure}

\end{itemize}

\begin{figure}
\vskip 0.2in
\begin{center}
\centerline{\includegraphics[width=1.1\columnwidth]{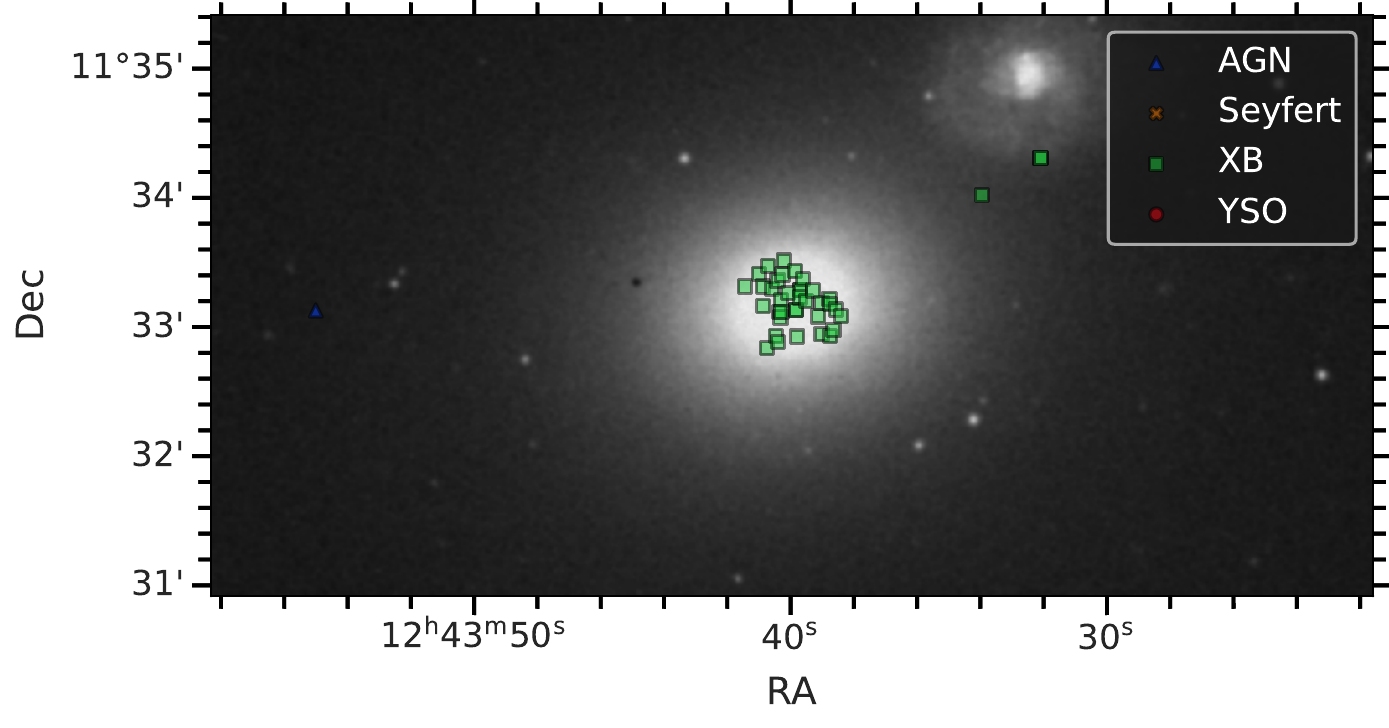}}
\caption{Previously unclassified sources in the region of \textit{NGC 4649} overlaid on a DSS NIR band image in grayscale. The symbols indicate the master classes assigned by our algorithm.}
\label{fig:ngc4649}
\end{center}
\vskip -0.2in
\end{figure}

We also validate membership for objects classified as X-ray binaries. Source 2CXO J193309.5+185902, for example, is classified as a HMXB by our method with just one detection and a probability of $95\%$. This is consistent with other works such as \cite{yang2022classifying}, in which the source is classified as a HMXB with a probability of $69 \pm 14\%$. More generally, we can look for \textbf{XB} in regions where such type of X-ray source is expected, such as the spiral arms of galaxies. We explored source aggregated classifications over the region of the M33 galaxy, in the outskirts of which sources 2CXO J013301.0+304043 and 2CXO J013324.4+304402 are located. Both sources are classified as \textit{\textbf{XB}}, in agreement with other machine-learning based classifications, such as \cite{tranin2022probabilistic}. Some sources are also missclassified as X-ray binaries, based on comparison with independent classification. Source 2CXO J053747.4-691019, for example, is confidently classified as a \textit{HMXB} by our algorithm. However, it has been classified as pulsar PSR J0537-6910 located in the Large Magellanic Cloud \citep{linwebbandbarret}. We note, however, that in some X-ray binaries the compact object can be a pulsar \citep{wijnands1998millisecond}.

Source 2CXO J031948.1+413046 is located at the heart of the Perseus galaxy cluster. It is classified very confidently (${>}0.9$ probability in all three detections) as a \textit{Seyfert\_2} galaxy candidate. Other sources in this core region have been studied and suggested as either Type 1.5 or Type 2 Seyferts, such as 3C 84 \cite{nustarperseus, cataloguequasars}. These specific galaxies typically exhibit a brightly illuminated core, predominantly at infrared wavelengths. However, the number of Seyfert 2 galaxies in the Perseus cluster is limited, with \citep{ngc3147} being one of the previously identified examples.  

A different test astrophysical environment is provided by NGC 4649 and its surroundings. This is an elliptical galaxy located in the Virgo cluster with a population of X-ray sources detected by \emph{Chandra}, the vast majority of which has been classified as Low-Mass X-ray Binaries \citep{Luo_2013, dabrusco2014}. This classification is based on their overall X-ray properties, density, and spatial coincidence with a rich population of Globular Clusters detected in Hubble-Space Telescope images \citep{strader2012deep}. All of the previously unclassified CSC sources near NGC~4649 for which we are providing a master class are assigned a \textit{XB} label by our pipeline, as shown in \autoref{fig:ngc4649}. The CSC source closest to the X-ray source identified by \cite{Luo_2013} as the AGN in the center of NGC4649 is 2CXO J124339.9+113309. As a previously classified source, it was not assigned a label in our classification, but it is located in the cluster 0, under the GinPair (Galaxy in Pair of Galaxies) category from SIMBAD. The next closest source (2CXO J124339.8+113307) is classified as an X-ray Binary in each of the two detections included in the CSC.

\section{Discussion}
\label{sec:discussion}

% For instance, this may facilitate the discovery of rapidly spinning magnetar-forming mergers that lack a gamma-ray counterpart.
The advent of all-sky, time-domain X-ray surveying missions such as \emph{eROSITA}, and the proposed capabilities of upcoming probe and flagship facilities such as the Advanced X-ray Imaging Satellite (AXIS), Arcus, and eventually Lynx, underline the importance of a robust method for the classification of  X-ray sources. This is important not only  because of the increasing volume of high-energy sources detected, but also for the design of multi-wavelength follow-up campaigns of particular objects of interest, in particular those that inform models of transient behavior and extreme accretion. Classification attempts are usually limited by two factors: the lack of a robust and balanced training set across X-ray classes, and the fact that a significant number of X-ray sources lack optical or infrared counterparts that could provide useful insights for the categorization of sources. Leveraging the information contained in the \emph{Chandra} Source Catalog, we have designed an unsupervised classification algorithm that acknowledges those limitations, with the scope of investigating to what extent the X-ray properties alone can provide relevant insights for the labeling of X-ray sources. 

We now discuss our findings, focusing on four major aspects: i) the performance of our method with respect to existing classification approaches and curated catalogs; ii) the implications for astrophysics of the relationship between specific classes and the distribution of X-ray properties; iii) the impact of multiple detections of a source on classification, and iv) the limitations of our method.

\subsection{Comparison to other methods and catalogs}
\label{section:discussion_1}
We compare our classification results with the following independent methods and compiled catalogs:

\begin{itemize}
    \item The training dataset described in \cite{yang2022classifying} in the context of supervised classification, containing $2947$ literature verified CSC X-ray sources classified as AGN, Cataclismic Variable (CV), high-mass star, HMXB, low-mass star, LMXB, neutron star, and YSO. (MUWCLASS TD)
    
    \item The catalog of confidently classified CSC sources presented from the \cite{yang2022classifying} pipeline, which contains $~31000$ CSC sources, with the same labels as the previous items. (MUWCLASS CCGCS)
    
    \item The machine learning classification approach of XMM-Newton 4XMM-DR10 catalog sources, presented in \cite{tranin2022probabilistic}. This method assigns four different labels: AGN, stars, X-ray binaries (XRBs), and cataclysmic variables (CVs). (4XMM-DR10 Classification)
    
    \item The catalog of $224168$ galaxies presented in \cite{toba2014luminosity} based on WISE and SDSS data. (WISE-SDSS Galaxies)
    
    \item The catalog of $121$ new redshift obscured AGNs of the Chandra Source Catalog presented in \cite{sicilian2022x} (Obscured AGNs).
\end{itemize}

\begin{table*}
\resizebox{2\columnwidth}{!}{\begin{tabular}{lllrllllllllll}
\toprule
Catalog name & Class & \# matches &  Precision & \textit{R} Precision & Recall\\
\midrule
MUWCLASS TD \citep{yang2022classifying} & AGN+Seyfert & 31 &  0.32  & 0.56 & 0.83 \\ 
& XB & 29  & 0.17 & 0.62 & 0.31 \\
& YSO & 56  & 0.18 & 0.74 & 0.83 \\
\midrule
MUWCLASS CCGCS \citep{yang2022classifying} & AGN+Seyfert & 922 &  0.94 & 0.98 & 0.62 \\ 
& XB & 627 & 0.05 & 0.06 & 0.51\\
& YSO & 255 & 0.36 & 0.51 & 0.72 \\
\midrule
4XMM-DR10 Classification \citep{tranin2022probabilistic} & AGN+Seyfert & 912 &  0.76 & 0.78 & 0.62 \\ 
& XB & 659 & 0.42 & 0.45& 0.42\\
\midrule
WISE-SDSS Galaxies \citep{toba2014luminosity} & AGN+Seyfert & 8 &  1 & 1 & 0.88 \\
\midrule
Obscured AGNs \citep{sicilian2022x} & AGN+Seyfert & 27 &  1 & 1 & 0.19 \\
\bottomrule
\end{tabular}
}
\caption{A summary of the comparison of our classification with previous work and catalogs. The number of matches corresponds to the number of sources belonging to our assigned class resulting from crossmatching the Chandra Source Catalog names or a matching criteria of ${\leq}1\arcsec$. }
\label{tab:summ_comparison}
\end{table*}

To perform the comparison, we cross-matched our catalog of  \textit{uniquely classified} sources with each of the reference catalogs, either using a coordinate radius (${\leq}1\arcsec$) or matching exact CSC names. Where appropriate, we merge similar classes into a single class prior to comparison. For example, we merge the \textit{\textbf{AGN}} and \textit{\textbf{Seyfert}} aggregated classes into a AGN+Seyfert class. We also merge our LMXB and HMXB classes into a single XB class. \autoref{tab:summ_comparison} summarizes the results of this comparison. The table lists for each case the number of matches, the precision, and the recall. The \emph{Precision} column refers to the precision when all labels in the reference catalog are included, whereas the \emph{R Precision} columns refers to the precision when only reference catalog labels that are in both catalogs are considered.

We note a significant increase in the precision when only the restricted set of labels is considered. In particular, 98\% of the objects that we classify as AGN+Seyfert actually belong to that class according to \citep{yang2022classifying}, with precision in other classes ranging from 6\% to about 80\%. The lowest precision is obtained for X-ray binaries, for which only 6\% of the objects that we classified as XBs actually belong to that class according to the reference catalog. The recall numbers tend to be more uniform across classes, with percentages ranging from 20\% to 88\%. For example, of all AGN+Seyfert objects in the \citep{yang2022classifying} training set, we correctly classify as such 83\% of them, the same proportion as for YSOs. The recall for the AGN+Seyfert class increases to 88\% when the WISE-SDSS catalog is considered. We observe the worst recall performance when trying to correctly classify obscured AGNs. From the \citep{sicilian2022x} catalog, we have only classified about 20\% of their objects as AGNs or Seyfert. 

In general, most classifications we labelled as AGN+Seyfert are accurate with respect to the MUWCLASS set, and we also recover a fair amount of objects actually belonging to that classes. X-ray binaries show comparatively lower precision and recall. We can recover about 50\% of them from control catalogs, but we tend to assign XB labels to objects that are classified in other categories (mostly AGN) in control catalogs. Spectral similarities between these two types of accreting objects is a likely reason for this degeneracy, which can be broken if the distance to the sources is known \citep{volonteri2017high}. A significant fraction of the objects that we classify as YSOs are classified as low-mass stars in the MUWCLASS training set. 62\% of the objects classified as stars in MUWCLASS, on the other hand, have YSO labels in SIMBAD.

% This is the first indication that additional properties might be helpful in distinguishing between these classes, a distinction that currently appears ambiguous within the X-ray property space.
We find $2006$ matching sources with the 4XMM-DR10 classification catalog. Almost 80\% of the sources that we classify as AGN or Seyfert galaxies agree with the \cite{tranin2022probabilistic} probabilistic classification, and of all AGNs and Seyfert galaxies in the 4XMM-DR10 catalog, we recover 62\% of them with the correct class. Again, X-ray binaries show lower precision and recall, likely due to confusion with supermassive accretors. Of the $2006$ matched sources, $435$ were categorized as \textbf{YSO} in our classification scheme. The majority of these sources were classified either as Star or XRB by the 4XMM-DR10 classification. We obtain high precision and recall for optically or IR-selected galaxies in \cite{toba2014luminosity}, although these are based on a very small number of matches.

\subsection{Astrophysical implications}

\subsubsection{Accretion-powered sources}

One remarkable result is that when we limit ourselves to the classes assigned by our algorithm, the confusion between small accretors and large accretors is not widespread. Such confusion would be expected on the basis of similar properties due to the accretion nature of the X-ray emission \citep{padovani2017active}. With certain degree of confidence, however, in our analysis we can distinguish X-ray binaries from AGNs, QSOs and Seyferts when only the X-ray properties are considered. As \autoref{fig:results_class_confusion_raw_grouped_out} shows,  large accretors are correctly classified as such in 75\% of the cases, and incorrectly classified as X-ray binaries only on 19\% of the cases. On the other hand, small accretors are correctly classified as X-ray binaries in 66\% of the cases, and incorrectly classified as large accretors in 20\% of the cases. 

The feature distribution maps of \autoref{fig:results_class_exp_hard} and \autoref{fig:results_class_exp_var} suggest that the differentiation is possible based on a wider range of hardness ratios in X-ray binaries compared to AGNs and Seyferts, as well as a slightly increased average hardness for X-ray binaries, in particular HMXBs. The broader range of hardness ratios can be interpreted in terms of the different spectral states of X-ray binaries, which occur in much shorter timescales than in AGNs \citep{remillard2006x}. Within the large accretors, \textbf{AGN} and \textbf{Seyfert} galaxies are often confused, which is likely the result of a nomenclature difference, Seyferts being a particular type of AGN \citep{peterson1997introduction}. We note, however, that Seyferts show more variability than AGNs, and Seyfert 2 galaxies in particular are more likely to be variable in the soft band, probably due to more obscuration due to their orientation with respect to the line of sight, in the context of the AGN unified model. QSOs in our sample, on the other hand, have tighter constrains in their hardness ratios and variabilities, consistent with the absence of spectral variability. In particular, they are less variable in the soft band with respect to AGNs and Seyferts. Stochastic variability in the thermal soft band of AGN spectra is usually related to disk instabilities, jet activity, or spottiness of infalling matter \citep{jovanovic2009x}, all of which might be more stable at the the high accretion rates seen in QSOs, provided that they do not have a blazar-like orientation. Hard band variability, on the other hand, relates to the hot corona \citep{ballantyne2020sustaining, petrucci2001testing, haardt1991two, galeev1979structured}, and appears more prominent in Seyfert 1 galaxies, for which the line of sight intersects the corona more directly \citep{soldi2014long}. Heavy absorption by the torus can also enhance both soft band variability the hardness ratios in Seyfert 2 galaxies due to preferential absorption of soft photons \citep{turner1997asca,risaliti1999distribution}. The results in \autoref{fig:results_class_exp_hard} and \autoref{fig:results_class_exp_var} show overall agreement with this unified model.

Compared to QSOs and AGNs, objects classified as X-ray binaries show a larger range of variability in the broad band, resembling the patterns observed in Seyfert galaxies, with a slightly increased broad band variability. Resolved non-thermal coronal X-ray emission is likely to be at the root of this similarity between systems of significantly different masses and physical scales. Recent studies have provided evidence of 
similar spectral state evolution in the luminosity-hardness diagram, related to the accretion processes in both XBs and AGNs  \citep{fernandez2021x}. No dusty torus is present in X-ray binaries, however. This might explain why in \autoref{fig:results_class_exp_var} both LMXBs and HMXBs tend to resemble more the variability pattern of Seyfert 1 galaxies, for which the line of sight intersects less obscuring material, and more of the corona. 

% Ask Jack?

We recognize that the incorporation of multi-wavelength counterparts can improve the separation between AGN-type sources and X-ray binaries, in particular if sources can be associated to optically derived redshifts, as has been shown for example in \citet{yang2022classifying}. Association with host galaxies can also help in breaking degeneracies between these two types of accretors. We point out, however, that only a fraction of X-ray sources have optical and IR counterparts. In particular, less than half of the CSC sources have at least one optical or IR counterpart when a cross-match is performed using the Gaia DR3, Legacy Survey DR10, PanSTARRS-1, and 2MASS catalogs\footnote{\url{https://cxc.cfa.harvard.edu/csc/csc_crossmatches.html}}. This highlights the importance of assessing the quality of the classification with only X-ray information is used.

%%%%%%%%MAKE SURE TO INCLUDE A SIMILAR DISCUSSION FOR YSOs
%%%%%%%%MAKE SURE TO INCLUDE A SIMILAR DISCUSSION FOR YSOs
%%%%%%%%MAKE SURE TO INCLUDE A SIMILAR DISCUSSION FOR YSOs
%%%%%%%%MAKE SURE TO INCLUDE A SIMILAR DISCUSSION FOR YSOs%%%%%%%%MAKE SURE TO INCLUDE A SIMILAR DISCUSSION FOR YSOs
%%%%%%%%MAKE SURE TO INCLUDE A SIMILAR DISCUSSION FOR YSOs
%%%%%%%%MAKE SURE TO INCLUDE A SIMILAR DISCUSSION FOR YSOs

%To differentiate these classes more effectively, we suggest the inclusion of additional properties, such as redshift, which could help to reduce their overlap, given the substantial difference in distances between these classes. This highlights an example where distance information proves to be a crucial discriminator. Moreover, we have noted that most previous studies on classification incorporate location as an input feature, a factor we have not considered in our methodology. The level of confusion between classes provides valuable insights into the degree to which X-ray data can distinguish between classes, and highlights potential additional information that could facilitate improved separation and, ultimately, more accurate classification.

\subsubsection{Young stars}

Objects classified as young stars have consistently higher variabilities, in particular in the broad band. T-Tauri stars are the only type of object classified by our pipeline for which the majority of examples have high variability both in the broand an soft bands. High levels of variability may be tied to irregularities in the accretion processes occurring within these young stellar objects \citep{testa2010x, preibisch2005origin}. In addition, it could be linked to strong magnetic fields, which could lead to prominent magnetic activity such as flares and coronal mass ejections \citep{testa2010x}. This high level of variability observed is consistent with our understanding of the the processes occurring at this stage of stellar evolution. While a similar level of magnetic activity is expected from Orion V* stars, we do not observe it in our sample. 

Objects classified as YSOs also show a bimodal distribution in their hardness ratios. Hard X-ray emission during coronal events might be partly responsible for this behavior, but degeneracy with other types of hard-emitters, such as HMXBs, can also be associated to the bimodal distribution. In fact, missclassification of other types of objects as YSO (the class with the highest classification accuracy in validation) is not uncommon, although lower that you would expect from the limited amount of information encoded in the X-rays. For example, source 2CXO J033829.0-352701, located at the heart of the weakly active galaxy NGC 1399 within the Fornax Cluster, was classified very confidently as a \textit{YSO} by our method, with a mean probability of 1.0 across six detections. It has been classified as an elliptical galaxy \citep{de1991third} with a Seyfert 2 AGN \citep{veron2006catalogue}. The source is soft ($\texttt{hard\_hs} < -0.5$) and shows remarkably low variability probability ($\texttt{var\_prob\_b} < 0.2$). While Seyfert 2 galaxies tend to be associated with harder spectra (the line of view intersects the obscuring torus that preferentially absorbs soft-X-ray photons), the specific orientation of the system with respect to the line of sight, as well as different contributions from scattering and reflection, can result in a softer spectrum.

We examined the properties of source detections with a \textit{YSO} label in SIMBAD, as well as those labelled as YSOs in our classification, and found that the density distributions of properties are consistent for both the reference set and the classified set. There are, however, some difference in certain regions of the parameter space that could lead to misclassifications, because the reference set is not fully representative of the population.

\subsection{Differences in the hard and soft voting classifiers}
Our catalog of \textit{ambiguously classified} objects comprises sources whose hard and soft classification disagree. That is, for these sources, the most common class among detections is not the same class with the highest average probability. These are cases in which there is ambiguity between two different classes of similar types, such as AGNs and Seyferts. We investigate some of these objects in order to gain additional astrophysical insight about how the algorithm asigns classes.

Source 2CXO J004228.2+411222 has $66$ detections and is classified as a \textit{LMXB} by the hard voting classifier, and as \textit{HMXB} by the soft voting classifier. Independent work has classified it as an X-ray binary/black hole candidate in M31 \citep{arnason2020identifying, barnard2014fifty, barnard2013chandra}. On the other hand, 2CXO J004246.9+411615, which has also been classified as an X-ray binary or black hole candidate in M31 \citep{barnard2014fifty, barnard2013chandra} and has 22 detections, is classified as a \textit{Seyfert\_1} by the hard voting classifier, and as a \textit{LMXB} by the soft voting classifier. In this latter case the mean probabilities are 0.22 for the \textit{Seyfert\_1} class and $0.23$ for the \textit{LMXB}) class. So, whereas in the case of 2CXO J004228.2+411222 the ambiguity can be interpreted in terms of a difference of spectral hardness between two types of X-ray binaries and our method still assigns a general class that agrees with previous literature, in the latter case the difference in probability between the two candidate classes is negligible, and the ambiguity is unlikely to be resolved without additional information, such as the redshift of the intrinsic luminosity of the source. Such differences between soft and hard classification should be considered when interpreting the results of our classification catalog.

The X-ray pulsar magnetar 2CXO J010043.0-721133 \citep{durant2005possible,durant2008search} is classified as an \textit{Orion\_V*} by the hard voting classifier, and as a \textit{YSO} by the soft voting classifier with $14$ detections. Both assigned classes fall in the aggregated class \textbf{YSO}. While the detection could be consistent with a pulsar with a fading optical counterpart \citep{durant2005possible, durant2008search}, the reason that it got assigned a YSO class is that the class "pulsar" is not among the classes that we have selected to be assigned to the classified objects. Young stars and pulsars might share some of their properties in the space of X-ray features.

% This is consistent with their expected prevalence in the Galactic plane and the youth of their star progenitors \citep{figer2005massive}. 

Finally, source 2CXO J020938.5-100847 is classified as a \textit{TTau*} by the hard voting classifier, and as an \textit{AGN} by the soft voting classifier. This source only has $2$ detections. Mean probabilities are $0.19$ for \textit{TTau*} and $0.44$ for \textit{AGN}, which favor the $AGN$ label. In fact, the source is located less than 2 arcseconds from the nuclear region of galaxy NGC 838 in the group HCG 16 \citep{o2014deep}, which has been spectroscopically classified as an active galaxy (LINER type). In this case, the mean probabilities are more informative of the true class of the objects. Overall, if the three aggregated classes (AGN+Seyfert, XB, YSO) are used, 46\% of the sources in the list of ambiguously classified sources show agreement between the two voting systems. This suggests that even when there is disagreement between the two voting systems for the fine classification between the labels in \autoref{fig:results_class_confusion_raw_out}, a significant fraction of these ambiguous sources can still be successfully classified in the more general classes of \autoref{fig:results_class_confusion_raw_grouped_out}. Taken all together, the probabilistic elements of our classification can provide insight into the nature of these objects, even in the absence of optical counterparts.

\subsection{Limitations and future work}
\label{sec:limits}

%In this subsection we explore additional limitations of the pipeline and results presented in this study. Furthermore, we discuss how these limitations could prompt future research, potentially enhancing both the performance and the impact of our findings.

We have demonstrated that a unsupervised approach to the classification of X-ray sources is possible even in the absence of an optical confirmation. Despite its applicability to X-ray catalogs, the GMM approach that we have adopted has a number of limitations that primarily stem from two factors: \emph{a)} the intrinsic degeneracy in the observational features of sources of different types, and \emph{b)} uncertainties in the knowledge base used (i.e. SIMBAD) to associate cluster membership to specific classes. 

Feature degeneracies can be mitigated by incorporating ancillary information about individual sources, such as environmental properties (galaxy host information, among others). While this is in principle possible for a fraction of X-ray sources, it remains unfeasible for X-ray sources without associated optical counterparts. Limitations on the knowledge base can be addressed, in principle, by constructing detailed training sets based on spectroscopic classification or other independent signatures. An example of this is the training set presented in \citet{yang2022classifying}. However, it is extremely difficult to construct such training sets while ensuring that they remain representative, balanced, and unbiased. This is both due to class imbalance and lack of independent classifications for X-ray sources in the majority of cases. Furthermore, we refrain from using coordinates as features due the CSC's skewed and non-homogeneous sky coverage, which could introduce biases. We therefore embrace these limitations as a challenge to push the limits of X-ray only unsupervised classification. In our experiments, we found our method to be robust against class imbalance, outperforming traditional classification algorithms. To evaluate the impact of class imbalance and compare our approach's performance with standard supervised methods, we detail a classification exercise using Support Vector Machines in online Appendix B.

%During data processing, we identified missing values (NaN) in some properties as a constraint. We addressed this by dropping these values. Usually, machine learning-focused studies opt for imputation or flagging missing points with extreme values so that the models can isolate them (e.g.,\cite{yang2022classifying}). A possible future development is to modify the clustering algorithm to take into account upper limits of those missing values, which are available in the Chandra Source Catalog. 

Gaussian Mixture Models are constrained by the assumption of Gaussian-distributed clusters, which does not hold for most distributions. Their parametric nature also restricts their ability to fit multidimensional data distributions in practical problems \citep{mallapragada2010non}. We used heuristic methods (such as the BIC and elbow methods) and identification of property distribution across multiple experiments to choose a number of Gaussian components that matches the overall behavior of the feature distributions. Alternative methods involve the use of Gaussian Mixtures with a Dirichlet process prior \citep{teh2010dirichlet, gorur2010dirichlet}, or other non-parametric proposals \citep{mallapragada2010non}. Also, spectral clustering is a generalization of classical clustering methods and is capable of identifying non-convex clusters \citep{hastie2009elements}.

The pipeline proposed here is similar to the \textit{cluster-then-label} semi-supervised method, wherein a clustering algorithm is first employed, followed by a supervised learner to label the unlabeled points in each cluster \citep{zhu2009introduction}. In our case, the Mahalanobis distance acts as the supervised learner, although each cluster can contain multiple classes, which justifies our probabilistic approach. A recurrent version of this model allows for the rapid classification of new detections, which would allow for the incorporation of the model to data processing pipelines of \emph{Chandra} or other observatories. Using our method, sources with diverging hard and soft voting classifications could be used to recognize objects with significant spectral or flux variability between observations, such as X-ray binaries transitioning between states. 

It should be noted that comparisons of probabilities between source detections in different clusters may not be informative for the purposes of comparing those two observations. That is because the probabilities presented in our results are relative to each cluster's individual space. For instance, the assigned probability of $100\%$ for source 2CXO J020938.4-100846 (as presented in Section \ref{section:results}) as an HMXB implies that within its specific cluster, the reference HMXB distribution is the closest to the source detection. 

%Another limitation of our approach is that it is possible that some sources do not belong to any of the classes that we selected as targets in reality. For example, we do not have a \textit{Star} class, which has been used in other machine learning related works (e.g., \cite{yang2022classifying}). We find in Section \ref{section:discussion_1} that most of the sources that we classify as \textit{YSO} are, as expected, part of the correspondent \textit{Star} class in other works.

We emphasize that our pipeline is not only easily reproducible but also adaptable to other catalog classification endeavors. We invite interested readers to explore the provided \texttt{GitHub} repository and implement proposed modifications. This could help enhance the performance of the work or enable exploration of other applications, such as the classification of nature-specific types of objects (e.g., \textit{YSO} and \textit{non-YSO} classification). 

\section{Summary and conclusions}
\label{sec:conclusions}
The automatic classification of X-ray sources is a necessary step in extracting astrophysical information from compiled source catalogs. Classification is useful for the study of individual objects, statistics for population studies, as well as for anomaly detection, i.e., the identification of new unexplored phenomena, including transients and spectrally extreme sources. This is true for \emph{Chandra} data, and for other X-ray missions that have compiled catalogs of X-ray sources and their properties, including \emph{XMM-Newton}, \emph{eROSITA}, etc. Despite the importance of this task, classification of astronomical X-ray sources remains challenging. This is due to mainly two issues: \emph{i)} a significant fraction of X-ray sources lack optical or infrared counterparts that could provide useful ancillary information, such as redshift; and \emph{ii)} the construction of reliable training sets that would allow automatic supervised learning classification is not trivial. These training sets are incomplete and unbalanced at best, due to the lack of spectroscopic confirmation for a significant number of sources.

Despite these difficulties, supervised classification of X-ray source catalogs has been attempted \citep{yang2022classifying, chen2023population, kumaran2023automated}. Such efforts rely on the ability to obtain multi-wavelength properties or observables for the sources to be classified, which is not always possible. In this work, we have developed an alternative methodology that employs unsupervised machine learning to provide probabilistic classes to Chandra Source Catalog sources. The method relies on clustering of X-ray properties such as hardness ratios and variability using Gaussian Mixtures. It suffers less from the lack of a reliable training set, as classes are assigned by association with cluster membership, as well as some topological considerations. Here are our main findings:

\begin{enumerate}

    \item As an alternative to directly learning from a representative set of labelled multi-wavelength training sources, we have demonstrated that it is possible to assign probabilistic classes to X-ray sources by first clustering them according to the distribution of X-ray properties, and then using a metric (the Mahalanobis distance) within each cluster to evaluate their probabilistic classes based on their distance to a much less representative set of independently classified sources.

    \item We provide a catalog of probabilistic classes for 8,756 sources, comprising a total of 14,507 detections. For each source, we provide a probability distribution over classes, as well as a measure of the uncertainty in its assigned class. We have validated our classification by using an internal validation set, and by comparing our classification with those performed using supervised approaches. 

    \item Our methodology is particularly successful at identifying emission from young stellar objects, with an 88\% overall accuracy. When compared with existing classification catalogs, our classification can reach a level agreement above 90\% in identifying AGNs and Seyfert galaxies.

   \item Our method is able to distinguish between small and large compact accretors (that is, X-ray binaries and active galactic nuclei) with more than 50\% confidence, despite the widespread assumption that such accreting systems can be very similar in their X-ray properties, and difficult to separate in two different classes in the absence of a redshift or distance measurement. This distinction is possible due to a wider range of hardness ratios in X-ray binaries compared to AGNs and Seyferts, along with a slightly higher average hardness for X-ray binaries,  in particular HMXBs.

   \item Compared to QSOs and AGNs, objects classified as X-ray binaries show a larger range of variability in the broad band, resembling the patterns observed in Seyfert galaxies, with a slightly increased broad band variability. Resolved non-thermal coronal X-ray emission is likely to be at the root of this similarity between systems of significantly different masses and physical scales.

   \item Differences in spectral variability and hardness ratios between objects classified as QSOs, Seyfert 1 and 2 galaxies, and other AGNs, are consistent with the unified AGN model. 
   
   \item Broad band variability stands as a key feature to differentiate XBs from AGNs, when relying solely on X-ray data. 

   \item Stellar variability due to coronal ejections in young stars is manifested in a distinct distribution of variabilities and hardness ratios in these objects, that informs their classification. Short-lived flares of hard X-ray photons are a likely cause of this distribution of the X-ray features.
    
    \item Our unsupervised method is similar to a semi-supervised method, with a probabilistic metric (e.g. the Mahalanobis distance) replacing a classical supervised classifier.

\end{enumerate}

Our complete results are accessible in the online supplementary material. Reproducibility is facilitated through the code available in our GitHub repository: \url{https://github.com/samuelperezdi/umlcaxs}. Additionally, we provide a notebook for classifying individual X-ray sources\footnote{\url{https://github.com/samuelperezdi/umlcaxs/blob/main/classify_your_source.ipynb}}. We offer a web app playground for direct data interaction: \url{https://umlcaxs-playground.streamlit.app/}. 
Here, users can access visualizations of source positions in the sky and their properties, along with respective classifications.

\section*{Acknowledgements}

We would like to express our gratitude to Hui Yang, Elias Krytsis, and Andreas Zezas for insightful comments, discussions and support. This research was made possible through the support from NASA/Chandra contract AR1-22004X "Machine Learning Discovery of X-Ray Transients in the Chandra Source Catalog 2.0" and the CXC Research Visitor Program. The provided funds facilitated the scientific visit of V. S. Pérez-Díaz to the Center for Astrophysics (CfA) $\mid$ Harvard \& Smithsonian, which allowed further exploration of the results and significant discussions. This visit was also possible thanks to Universidad del Rosario, which supported travel expenses through the IV-YIG001 FONDO DE INVESTIGACIÓN POR PRODUCCIÓN ACADÉMICA. R.D'A. is supported by NASA contract NAS8-03060 (Chandra X-ray Center). We acknowledge the use of Chandra Source Catalog 2.0 data. This work was developed under the scope of AstroAI\footnote{\url{http://astroai.cfa.harvard.edu/}}, a new center dedicated to the development of artificial intelligence to enable next generation astrophysics at the CfA. For the purpose of open access, we have applied a Creative Commons Attribution (CC BY) license to any Author Accepted Manuscript version resulting from this submission.

%%%%%%%%%%%%%%%%%%%%%%%%%%%%%%%%%%%%%%%%%%%%%%%%%%
\section*{Data Availability}

The data supporting this article can be found within the article itself and its associated online supplemental content.

Supplementary files are available at MNRAS online.

appendix.pdf


% \nocite{chandra_20_book}
% \nocite{yangetal}
% \nocite{Evans2010}
\bibliographystyle{mnras}
\bibliography{references} % if your bibtex file is called example.bib

%\begin{thebibliography}{99}
%\end{thebibliography}

%%%%%%%%%%%%%%%%%%%%%%%%%%%%%%%%%%%%%%%%%%%%%%%%%%

%%%%%%%%%%%%%%%%% APPENDICES %%%%%%%%%%%%%%%%%%%%%
\appendix

\section{Machine readable tables}

\begin{landscape}
\begin{table}
\resizebox{1.3\textwidth}{!}{\begin{tabular}{lllrllllllllll}
\toprule
{} & agg\_master\_class & master\_class &  detection\_count &          QSO &                   AGN &             Seyfert\_1 &    Seyfert\_2 &                  HMXB &         LMXB &           XB &                   YSO &        TTau* &     Orion\_V* \\
name                  &                  &              &                  &              &                       &                       &              &                       &              &              &                       &              &              \\
\midrule
2CXO J004231.1+411621 &               XB &         HMXB &               97 &  0.056±0.152 &           0.156±0.209 &           0.079±0.151 &   0.04±0.106 &  \textbf{0.241±0.215} &  0.213±0.228 &  0.142±0.222 &           0.024±0.066 &  0.004±0.016 &  0.044±0.164 \\
2CXO J004248.5+411521 &               XB &         HMXB &               93 &  0.039±0.107 &           0.113±0.175 &           0.121±0.174 &  0.091±0.193 &  \textbf{0.274±0.254} &  0.159±0.185 &  0.169±0.242 &           0.012±0.031 &  0.003±0.014 &  0.019±0.099 \\
2CXO J123049.3+122328 &          Seyfert &    Seyfert\_1 &               85 &  0.113±0.249 &           0.138±0.241 &  \textbf{0.467±0.397} &  0.102±0.251 &           0.021±0.106 &  0.017±0.115 &  0.023±0.121 &           0.035±0.128 &  0.024±0.106 &  0.059±0.184 \\
2CXO J004254.9+411603 &               XB &         HMXB &               84 &  0.053±0.165 &           0.096±0.147 &           0.065±0.119 &  0.093±0.219 &  \textbf{0.309±0.256} &  0.179±0.186 &  0.152±0.201 &           0.032±0.125 &  0.008±0.041 &  0.013±0.035 \\
2CXO J004232.0+411314 &               XB &         HMXB &               80 &  0.014±0.081 &           0.059±0.142 &           0.153±0.262 &  0.257±0.348 &  \textbf{0.335±0.367} &  0.094±0.192 &   0.07±0.179 &           0.006±0.032 &    0.0±0.001 &  0.012±0.105 \\
2CXO J004247.1+411628 &               XB &         HMXB &               71 &  0.054±0.177 &           0.078±0.142 &           0.076±0.157 &  0.071±0.185 &  \textbf{0.284±0.263} &  0.179±0.234 &  0.161±0.235 &           0.039±0.119 &  0.002±0.006 &  0.057±0.196 \\
2CXO J004213.1+411836 &              YSO &          YSO &               71 &    0.0±0.002 &               0.0±0.0 &               0.0±0.0 &    0.0±0.002 &           0.003±0.014 &  0.027±0.157 &      0.0±0.0 &  \textbf{0.505±0.374} &   0.11±0.207 &   0.355±0.35 \\
2CXO J004257.9+411104 &              AGN &          AGN &               60 &  0.062±0.106 &  \textbf{0.194±0.238} &            0.097±0.16 &  0.101±0.227 &           0.183±0.233 &  0.109±0.173 &  0.141±0.234 &           0.028±0.098 &  0.029±0.097 &  0.056±0.177 \\
2CXO J123049.4+122327 &              AGN &          AGN &               57 &  0.229±0.279 &  \textbf{0.267±0.344} &           0.188±0.249 &  0.096±0.244 &           0.015±0.096 &  0.035±0.184 &    0.0±0.003 &           0.121±0.323 &  0.009±0.067 &  0.039±0.172 \\
2CXO J123048.6+122332 &          Seyfert &    Seyfert\_1 &               53 &  0.109±0.251 &           0.242±0.361 &  \textbf{0.315±0.409} &  0.035±0.128 &           0.074±0.219 &  0.015±0.102 &  0.004±0.028 &           0.037±0.122 &  0.066±0.214 &  0.104±0.251 \\
\bottomrule
\end{tabular}
}
\caption{A sample of the \textit{uniquely classified} classification table, available in the online supplementary material and the paper's GitHub repository as \texttt{uniquely\_classified.csv}. Mean probabilities and standard deviations are highlighted in bold text for the master class. Note that additional property columns are not included in this sample.}
\label{A:table_confident}
\end{table}

\begin{table}
\resizebox{1.3\textwidth}{!}{\begin{tabular}{lllrllllllllll}
\toprule
{} & hard\_master\_class & soft\_master\_class &  detection\_count &          QSO &          AGN &    Seyfert\_1 &    Seyfert\_2 &         HMXB &         LMXB &           XB &          YSO &        TTau* &     Orion\_V* \\
name                   &                   &                   &                  &              &              &              &              &              &              &              &              &              &              \\
\midrule
2CXO J004228.2+411222  &              LMXB &              HMXB &               66 &  0.057±0.136 &  0.113±0.181 &  0.128±0.234 &  0.095±0.217 &  0.217±0.238 &  0.171±0.211 &   0.167±0.25 &  0.006±0.017 &  0.004±0.018 &   0.043±0.17 \\
2CXO J004246.9+411615  &         Seyfert\_1 &              LMXB &               22 &  0.074±0.143 &  0.063±0.088 &  0.219±0.258 &    0.118±0.2 &  0.125±0.148 &  0.226±0.236 &  0.115±0.181 &  0.005±0.013 &  0.003±0.013 &  0.051±0.203 \\
2CXO J004244.3+411608A &         Seyfert\_1 &               QSO &               15 &   0.24±0.297 &   0.09±0.126 &  0.162±0.203 &   0.092±0.25 &   0.06±0.125 &  0.027±0.056 &   0.012±0.03 &  0.155±0.284 &  0.014±0.041 &  0.147±0.277 \\
2CXO J004244.3+411608  &         Seyfert\_1 &               QSO &               15 &   0.24±0.297 &   0.09±0.126 &  0.162±0.203 &   0.092±0.25 &   0.06±0.125 &  0.027±0.056 &   0.012±0.03 &  0.155±0.284 &  0.014±0.041 &  0.147±0.277 \\
2CXO J010043.0-721133  &          Orion\_V* &               YSO &               14 &      0.0±0.0 &      0.0±0.0 &      0.0±0.0 &      0.0±0.0 &  0.009±0.022 &   0.121±0.29 &      0.0±0.0 &  0.423±0.408 &  0.035±0.101 &  0.412±0.379 \\
2CXO J004242.4+411553  &                XB &              HMXB &               13 &  0.005±0.007 &  0.068±0.115 &  0.135±0.137 &  0.108±0.188 &  0.277±0.256 &  0.153±0.154 &  0.247±0.269 &  0.001±0.002 &      0.0±0.0 &  0.006±0.022 \\
2CXO J063354.3+174614  &               AGN &         Seyfert\_1 &               11 &  0.221±0.197 &   0.348±0.29 &   0.419±0.33 &  0.011±0.031 &  0.001±0.002 &      0.0±0.0 &      0.0±0.0 &      0.0±0.0 &      0.0±0.0 &      0.0±0.0 \\
2CXO J133656.6-294912  &              LMXB &              HMXB &               11 &  0.001±0.002 &  0.034±0.041 &   0.01±0.016 &  0.022±0.062 &   0.373±0.31 &   0.361±0.24 &  0.106±0.127 &  0.001±0.002 &      0.0±0.0 &  0.091±0.287 \\
2CXO J061538.9-574204  &               YSO &              HMXB &               10 &      0.0±0.0 &      0.0±0.0 &      0.0±0.0 &      0.0±0.0 &     0.4±0.49 &  0.216±0.395 &      0.0±0.0 &  0.384±0.472 &      0.0±0.0 &    0.0±0.001 \\
2CXO J053816.3-692331  &          Orion\_V* &               YSO &               10 &      0.0±0.0 &      0.0±0.0 &      0.0±0.0 &      0.0±0.0 &    0.0±0.001 &      0.0±0.0 &      0.0±0.0 &  0.505±0.426 &      0.0±0.0 &  0.495±0.426 \\
\bottomrule
\end{tabular}

}
\caption{A sample of the \textit{ambiguous}  table, available in the online supplementary material and the paper's GitHub repository as \texttt{ambiguous\_classification.csv}. \textit{Hard} and \textit{soft} master classes are provided. Note that additional property columns are not included in this sample.}
\label{A:table_confused}
\end{table}
\end{landscape}

% Don't change these lines
\bsp	% typesetting comment
\label{lastpage}
\end{document}